\documentclass[aps,prd,nofootinbib,showpacs,preprintnumbers]{revtex4}
\usepackage[dvips]{epsfig}
\usepackage{amssymb}
\usepackage{verbatim}
\usepackage{psfrag}
\newcommand{\beqn} {\begin{equation}}
\newcommand{\eqn} {\end{equation}}
\newcommand{\hm}{\hat{m}}

\def \beq{\begin{equation}}
\def \eeq{\end{equation}}
\def \bea{\begin{eqnarray}}
\def \eea{\end{eqnarray}}

\def \bet0{\beta_0}
\def \bet1{\beta_1}
\def \simgt{\,\rlap{\lower 7.5 pt\hbox{$\mathchar \sim$}}\raise 3 pt \hbox{$>$}\,}
\def \simlt{\,\rlap{\lower 7.5 pt\hbox{$\mathchar \sim$}}\raise 3 pt \hbox{$<$}\,}

\def\lsim{\raise0.3ex\hbox{$<$\kern-0.75em\raise-1.1ex\hbox{$\sim$}}}
\def\gsim{\raise0.3ex\hbox{$>$\kern-0.75em\raise-1.1ex\hbox{$\sim$}}}

\newcommand{\VEV}[1]{\left\langle #1\right\rangle}
\newcommand{\pbp}{\langle \bar \psi \psi \rangle}
\newcommand{\pbdmdup}{\left\langle \bar \psi \frac{dM}{du_0} \psi \right\rangle}
\newcommand{\maths}[1]{$#1$}
\newcommand{\link}[2]{  U_{ #1 , #2}  }
\newcommand{\linkdagger}[2]{  U_{ #1 , #2}^\dagger  }
\newcommand{\linkdaggerfat}[2]{  U_{ #1 , #2}^{{\rm fat} \dagger}  }
\newcommand{\kronecker}[2]{  \delta_{#1 \; #2}  }

\newcommand{\apart}[3]{      
\sum_{#3} \eta_{#3}(#1)\Big( \link{#1}{#3}^{\rm fat} \kronecker{#1}{#2-\hat{#3}} 
                -\linkdaggerfat{#1-\hat{#3}}{#3} 
                                        \kronecker{#1}{#2+\hat{#3}} \Big) }

\newcommand{\bparta}[3]{
\sum_{#3} \eta_{#3}(#1)\Big( \link{#1}{#3} \link{#1+\hat{#3}}{#3} \link{#1+2 \hat{#3}}{#3}
\kronecker{#1}{#2- 3 \hat{#3}}
- \linkdagger{#1-\hat{#3}}{#3} \linkdagger{#1-2\hat{#3}}{#3}
\linkdagger{#1-3 \hat{#3}}{#3} \kronecker{#1}{#2+ 3 \hat{#3}} \Big) }

\newcommand{\bpartb}[4]{
\sum_{#3} \eta_{#3}(#1)\sum_{#4\ne#3}\bigg[ \Big( \link{#1}{#3} \link{#1+\hat{#3}}{#4} \link{#1+\hat{#3} + \hat{#4}}{#4} \kronecker{#1}{#2-\hat{#3}-2\hat{#4}}
- \linkdagger{#1-\hat{#4}}{#4} \linkdagger{#1-2\hat{#4}}{#4} 
 \linkdagger{#1- \hat{#3} - 2\hat{#4}}{#3} \kronecker{#1}{#2+\hat{#3}+2\hat{#4}}\Big)\\
&&~~~+ \Big( \link{#1}{#4} \link{#1+\hat{#4}}{#4} \link{#1+
2\hat{#4}}{#3} \kronecker{#1}{#2-\hat{#3}-2\hat{#4}} 
 - \linkdagger{#1-\hat{#3}}{#3} \linkdagger{#1-\hat{#3}-\hat{#4}}{#4} 
     \linkdagger{#1- \hat{#3} - 2\hat{#4}}{#4}\kronecker{#1}{#2+\hat{#3}+2\hat{#4}}\Big)\\
&&~~~+ \Big( \linkdagger{#1-\hat{#4}}{#4} \linkdagger{#1-2\hat{#4}}{#4}
\link{#1-2\hat{#4}}{#3} \kronecker{#1}{#2-\hat{#3}+2\hat{#4}}
 - \linkdagger{#1-\hat{#3}}{#3} \link{#1-\hat{#3}}{#4} \link{#1- \hat{#3} +
\hat{#4}}{#4} \kronecker{#1}{#2+\hat{#3}-2\hat{#4}}\Big)\\
&&~~~+ \Big( \link{#1}{#3} \linkdagger{#1+\hat{#3}-\hat{#4}}{#4}
\linkdagger{#1+\hat{#3}-2\hat{#4}}{#3} \kronecker{#1}{#2-\hat{#3}+2\hat{#4}}
 - \link{#1}{#4} \link{#1+\hat{#4}}{#4} \linkdagger{#1- \hat{#3} +
2\hat{#4}}{#3} \kronecker{#1}{#2+\hat{#3}-2\hat{#4}}           
          \Big) \bigg] }

\begin{document}
\title{Equation of state and QCD transition at finite temperature}
\author{A. Bazavov$^a$,
T. Bhattacharya$^{\rm b}$, M. Cheng$^{\rm c}$, N.H. Christ$^{\rm d}$,
C. DeTar$^{\rm e}$, S. Ejiri$^{\rm h}$,\\ 
Steven Gottlieb$^{\rm f}$, R. Gupta$^{\rm b}$, 
U.M. Heller$^{\rm g}$, K. Huebner$^{\rm h}$,  
C. Jung$^{\rm h}$, F. Karsch$^{\rm h,i}$,\\ 
E. Laermann$^{\rm i}$, L. Levkova$^{\rm e}$, C. Miao$^{\rm h}$,
R.D. Mawhinney$^{\rm d}$, P. Petreczky$^{\rm h,j}$, C. Schmidt$^{\rm i}$, \\
R.A. Soltz$^{\rm c}$, W. Soeldner$^{\rm k}$, 
R. Sugar$^{\rm l}$, D. Toussaint$^{\rm a}$ and P. Vranas$^{\rm c}$ 
}
\affiliation{
$^{\rm a}$ Physics Department, University of Arizona, Tucson, AZ 85721, USA\\
$^{\rm b}$ Theoretical Division, Los Alamos National Laboratory, Los Alamos, NM 87545, USA\\
$^{\rm c}$ Physics Division, Lawrence Livermore National Laboratory, Livermore CA 94550, USA\\
$^{\rm d}$ Physics Department,Columbia University, New York, NY 10027, USA\\
$^{\rm e}$ Physics Department, University of Utah, Salt Lake City, UT 84112, USA \\
$^{\rm f}$ Physics Department, Indiana University, Bloomington, IN 47405, USA\\
$^{\rm g}$ American Physical Society, One Research Road, Ridge, NY 11961, USA\\ 
$^{\rm h}$ Physics Department, Brookhaven National Laboratory,Upton, NY 11973, USA \\
$^{\rm i}$ Fakult\"at f\"ur Physik, Universit\"at Bielefeld, D-33615 Bielefeld, Germany\\
$^{\rm j}$ RIKEN-BNL Research Center, Brookhaven National Laboratory, Upton, NY 11973, USA \\
$^{\rm k}$ Gesellschaft f\"ur Schwerionenforschung, Planckstr.~1, D-64291 Darmstadt, Germany \\
$^{\rm l}$ Physics Department, University of California, Santa Barbara, CA 93106, USA\\
}

\begin{abstract}
We calculate the equation of state in 2+1 flavor QCD at finite temperature
with physical strange quark mass and almost physical light
quark masses using lattices with temporal extent $N_{\tau}=8$. Calculations
have been performed with two different improved staggered fermion actions,
the asqtad and p4 actions. 
Overall, we find good agreement between results obtained with these 
two $O(a^2)$ improved staggered fermion discretization schemes.
A comparison with earlier calculations on 
coarser lattices is performed to quantify systematic errors
in current studies of the equation of state. 
We also present results for observables that are sensitive to deconfining and 
chiral aspects of the QCD transition on $N_\tau=6$ and $8$ lattices. We find that
deconfinement and chiral symmetry restoration happen in the same narrow 
temperature interval.
In an Appendix we present a simple parametrization of the equation of state that 
can easily be used in hydrodynamic model calculations. In this parametrization we 
also incorporated an estimate of current uncertainties in the lattice calculations 
which arise from cutoff and quark mass effects. We estimate these systematic effects 
to be about $10$~MeV. 
\end{abstract}
\maketitle

\section{Introduction}

Determining the equation of state (EoS) of hot, strongly interacting matter is
one of the major challenges of strong interaction physics.  The QCD EoS provides
a fundamental characterization of finite temperature QCD and is a critical input
to the hydrodynamic modeling of the expansion of dense matter formed in heavy ion
collisions. In particular, 
the interpretation of recent results from RHIC on jet quenching, hydrodynamic 
flow, and charmonium production \cite{RHIC} rely on an accurate determination
of the energy density and pressure as well as an understanding of both the 
deconfinement and chiral transitions.  

For vanishing chemical potential, which is appropriate for 
experiments at RHIC and LHC, lattice calculations of the
EoS \cite{aoki,milc_eos,rbcBIeos} as well as the transition temperature
\cite{milc04,tc06,aoki_Tc} can be performed with an almost realistic
quark mass spectrum. In addition, calculations at different values of the lattice cutoff
allow for a systematic analysis of discretization errors and will soon
lead to a controlled continuum extrapolation of the EoS
with physical quark masses.

Studies of QCD thermodynamics are most advanced in lattice regularization
schemes that use staggered fermions. In this case, improved actions have been
developed that eliminate ${\cal O}(a^2)$ discretization errors efficiently 
in the calculation of bulk thermodynamic observables at high temperature as well
as at nonvanishing chemical potential \cite{Heller,Hegde}. At finite temperature,
these cutoff effects are controlled by the temporal extent $N_\tau$ of the 
lattice as this fixes the cutoff in units of the temperature, $aT=1/N_\tau$.
At tree level, which
is relevant for the approach to the infinite temperature limit, the 
asqtad \cite{asqtad,Orginos} and p4 \cite{p4,Heller} discretization schemes have been found to give rise to only small
deviations from the asymptotic ideal gas limit already on lattices
with temporal extent $N_\tau =6$. By $N_\tau = 8$, the deviations from the continuum 
Stefan-Boltzmann value are at the 1\% level~\cite{Heller}. At moderate 
values of the temperature, one expects that genuine 
nonperturbative effects will contribute to the cutoff dependence and, moreover,
as the relevant degrees of freedom change from partonic at high temperature
to hadronic at low temperature, other cutoff effects may become important.
Most notably in the case of staggered fermions is the explicit breaking 
of staggered flavor (taste) symmetry that leads to ${\cal O}(a^2)$ distortion of the hadron 
spectrum and will influence the thermodynamics in the confined phase. 

To judge the importance of different effects that contribute to the cutoff
dependence of thermodynamic observables, we have performed calculations with
two different staggered fermion actions which deal with these systematic 
effects in different ways. The asqtad and p4 actions combined with a 
Symanzik improved gauge action eliminate ${\cal O}((aT)^2)$ errors in 
thermodynamic observables at tree level. The asqtad action goes beyond
tree-level improvement through the introduction of nonperturbatively 
determined tadpole coefficients \cite{milc_eos}. In the gauge part of 
the action, this also means that in addition to the planar six link loop 
the nonplanar `parallelogram' has been introduced. Furthermore, both 
actions use so-called fat links to reduce the influence of taste symmetry
breaking terms inherent in staggered discretization schemes at nonzero 
values of the lattice spacing. The asqtad and p4 schemes also differ
in the way fat-links are introduced. 
In the asqtad action, fat link coefficients have been adjusted
so that tree-level coupling to all hard gluons has been 
suppressed without introducing further ${\cal O}(a^2)$ errors. 
The p4 action, on the other hand, only uses a simple three-link staple 
for fattening.

In this paper, we report on detailed calculations of the thermodynamics of strongly
interacting elementary particles performed in lattice-regularized QCD
with a physical value of the strange quark mass and with two degenerate 
light quark masses being one tenth of the strange quark mass. To study the 
quark mass dependence of the thermodynamic quantities,
we have also performed calculations at a larger value of the light quark mass
corresponding to one fifth of the strange quark mass. Combining with 
past results on QCD thermodynamics at vanishing as well as nonvanishing
values of the chemical potential on lattices with larger lattice spacing
allows for an analysis of cutoff effects within both discretization schemes.
In fact, a major obstacle to
quantifying cutoff effects in studies of the QCD equation of state is
that they arise from different sources which are strongly
temperature dependent, and their relative importance changes with
temperature. This makes it difficult to deal
with them in a unique way and make a direct comparison between results
obtained within different discretization schemes. It is, therefore, very important to 
understand and control systematic errors reliably.

In the next section, we start with a discussion of the basic 
setup for the calculation of the equation of state using the ${\cal O}(a^2)$
improved asqtad and p4 actions. We proceed with a presentation of results
for the trace anomaly that characterizes deviations from the conformal limit,
in which the energy density ($\epsilon$) equals three times the pressure ($p$).
In Sec.~III, we give results on several bulk thermodynamic observables
that can be derived from the trace anomaly $(\epsilon -3p)/T^4$ using 
standard thermodynamic relations. In Sec.~IV, we analyze the temperature
dependence of quark number susceptibilities, chiral condensates and the 
Polyakov loop expectation value and discuss in terms of them deconfining 
and symmetry restoring features of the QCD transition. We give our conclusions
in Sec.~V.  Furthermore, we give a coherent discussion of calculations
of the QCD equation of state with the asqtad and p4 actions in Appendix A.
Appendix B summarizes results for renormalization constants needed to
calculate the renormalized Polyakov loop expectation value with the asqtad
action.  In Appendix C we provide a parametrization for the equation of state
suitable for application to hydrodynamic modeling of heavy ion collisions.
All numerical results needed to calculate the bulk thermodynamic
observables presented in this paper are given in Appendix D. 

\section{Basic input into the calculation of the QCD Equation of State:
Trace Anomaly}

\subsection{Calculational setup}

This publication reports on a detailed study of the EoS of (2+1)-flavor QCD
on lattices with temporal extent $N_\tau =8$. It extends earlier studies 
performed with the asqtad and p4 actions on lattices with temporal
extent $N_\tau=4$ and $6$ \cite{milc_eos,rbcBIeos}. The calculational
framework for the analysis of the equation of state, the 
thermodynamic quantities that need to be calculated and the 
dependence of various parameters that appear 
in the gauge and fermion actions on the gauge coupling $\beta = 6/g^2$
has been discussed in these previous publications. It is, however, cumbersome 
to collect from the earlier publications all the 
information needed to follow the discussion given here, as the
calculational set-up and the specific observables that need to be 
evaluated differ somewhat between the tadpole improved asqtad action \cite{milc_eos} and the
tree level improved p4 action \cite{rbcBIeos}. In Appendix A we, therefore, give a coherent 
discussion of the various calculations performed with the asqtad and 
p4 actions, summarize the necessary theoretical background provided
previously in the literature and unify the different 
notations and normalizations used in the past by different groups working with
different staggered discretization schemes. In Appendix D, we give 
details of the simulation parameters and the statistics collected in each
of these calculations.

Most of the finite temperature calculations presented here have been performed 
on lattices of size
$32^3\times 8$ using the RHMC algorithm \cite{rhmc}. We combine these results with 
earlier calculations at $N_\tau=6$. For 
the asqtad action we use both previous results using the inexact R~algorithm 
\cite{milc_eos} and new RHMC calculations on lattices of 
size $32^3\times 6$, while the p4 results have all been obtained using the RHMC algorithm.
For each finite temperature calculation that entered our analysis of the
equation of state, a corresponding `zero temperature' calculation has 
been performed, mostly on lattices of size $32^4$, at the same value of the gauge 
coupling and for the same set of bare quark mass values, {\it i.e.}, at the same
value of the cutoff.

Following earlier calculations, we use
a strange quark mass that is close to its physical value and two degenerate
light quark masses that are chosen to be one tenth of the strange quark mass.
This choice for the light quark masses corresponds to a light pseudo-scalar 
Goldstone mass of about $220$~MeV\footnote{In the staggered fermion formulation
only one of the pseudo-scalar states has a mass vanishing in the chiral limit. 
The other states have masses that are of ${\cal O}(a^2)$ bigger and vanish 
only in the continuum limit. At cutoff values corresponding to the 
transition region of our calculations on $N_\tau=8$ lattices, these non-Goldstone
masses are of the order of 400~MeV for calculations with the asqtad
action and about 500~MeV with the p4 action.}.

All calculations have been performed on a line of constant physics (LCP), 
{\it i.e.}, as the temperature is increased the bare quark
masses have been adjusted such that the values of hadron masses in physical 
units, evaluated at zero temperature, stay approximately constant.
In practice, the LCP has been determined through the calculation of 
strange ($m_K$ or $m_{\bar{s}s}$) and nonstrange ($m_\pi$) meson masses in 
units of scales $r_n$ that characterize the shape of the static quark 
potential,
\begin{equation}
\left( r^2 \frac{{\rm d}V_{\bar{q}q}(r)}{{\rm d}r} \right)_{r=r_0} =
1.65 \;\; , \;\;
\left( r^2 \frac{{\rm d}V_{\bar{q}q}(r)}{{\rm d}r} \right)_{r=r_1} =
1.0  \;\; .
\label{r0r1}
\end{equation}

A major concern when comparing 
calculations performed with two different discretizations of the 
QCD action is quantifying systematic errors. 
A natural way to make such a comparison is to simulate the two actions with
a choice of parameters that give the same cutoff when expressed
in physical units. 
The most extensive calculations done by us with the two actions are of the 
parameters $\hat{r}_n\equiv r_n/a$ that define the shape of the heavy quark 
potential. We therefore determine the cutoff scale and define a common
temperature scale in units of $r_n$, {\it i.e.}, $r_nT\equiv r_n/a N_\tau$.
Note that for the comparison of results obtained with different actions, 
an accurate value of $r_n$ in physical units (1/MeV) is not necessary as 
only $r_nT$ is needed.

The ratio $r_0/r_1$ has been determined in the two 
discretization schemes consistently, as shown by the results 
$r_0/r_1 = 1.4636(60)$ (p4 \cite{rbcBIeos}) and $1.474 (7)(18)$
(asqtad \cite{asqtad_pot}).
We emphasize that these 
determinations of temperature scale based on 
parameters $\hat{r}_0$ and $\hat{r}_1$ were performed prior to the current 
combined analysis of thermodynamics with both the asqtad and p4 actions. 
Furthermore, this was done in completely
independent calculations using data analysis strategies and fitting routines
that have also been developed independently within the MILC \cite{milc_eos} and 
RBC-Bielefeld \cite{rbcBIeos} collaborations. 

To determine scales $r_0$ and $r_1$ in physical
units (MeV) we have related them to properties of
the bottomonium spectrum. As our final input, we use the value $r_0=0.469(7)$~fm 
determined from the $\Upsilon (2S-1S)$ splitting \cite{lepage,gray}
in calculations with the asqtad action. The same calculations show 
that the lattice scales from $r_0$ and $r_1$ 
are consistent with calculations
of the pion and kaon decay constants as well as mass splittings between
light hadronic states after extrapolations to the continuum limit and physical quark masses 
and agree with experimental results within errors of 3\%. 
Note that all these observables, sometimes called gold-plated 
observables \cite{gold_plated}, have been calculated within the
same discretization scheme at identical values of the cutoff as used 
in the finite temperature studies reported here. This  
consistency gives us 
confidence in using scales extracted from the heavy quark potential
for both extrapolating results to the continuum limit and for converting them to 
physical units. The scale $r_0 T$ is shown on top of the figures.

In Table~\ref{tab:LCP}, we summarize the masses that characterize
the LCP used for calculations with the p4 and asqtad actions.
We find that the lines of constant physics are similar
in both calculations, but differ in details. In particular, the strange
pseudo-scalar mass on the LCP used for calculations with the asqtad action 
is 15\% larger than the one used in the calculation with the p4-action. 
As the LCPs for asqtad and p4 actions had been fixed prior to this 
work in calculations on coarser lattices, we found it reasonable to 
stay with this convention rather than re-adjusting the choice of LCPs
for this work. This makes the comparison of cutoff effects within a 
given discretization scheme easier. The difference in LCPs, however, 
should be kept in mind when comparing results obtained with different actions.

\begin{table}[t]
\begin{center}
\vspace{0.3cm}
\begin{tabular}{|c|c|c|}
\hline
~ & p4-action & asqtad action \\
\hline
$m_{\bar{s}s} r_0$ &  1.59(5) &  1.83(6) \\
$m_{\bar{s}s}/m_K$ &  1.33(1) & 1.33(2) \\
$m_{\pi}/m_K$ &  0.435(2) &  0.437(3) \\
\hline
\end{tabular}
\end{center}
\caption{The strange pseudo-scalar mass
$m_{\bar{s}s} \equiv \sqrt{2m_K^2 - m_\pi^2}$ in units of $r_0$ and ratios 
of meson masses that characterize lines of constant physics for a fixed 
ratio of light and strange quark masses, $m_s/m_q = 10$. The errors are not 
statistical.  They represent a range of values over the lines of constant 
physics.
}
\label{tab:LCP}
\end{table}

The statistics and details of the data needed to calculate the basic thermodynamic
quantity, the trace anomaly, are summarized in Appendix \ref{data_summary}.

\subsection{The trace anomaly}
\label{sec:trace_anomaly}

Along the LCP, {\it i.e.}, for quark masses that are constant in physical 
units, and for sufficiently large volumes, temperature is the only
intensive parameter controlling the thermodynamics. Consequently, in our 
calculations there is only one independent bulk thermodynamic observable that needs
to be calculated. All other thermodynamic quantities are then obtained as appropriate
derivatives of the QCD partition functions with respect to the temperature and by 
using standard thermodynamic relations.

The quantity most convenient to calculate on the lattice is 
the trace anomaly in units of the fourth power of the
temperature $\Theta^{\mu\mu}/T^4$. This is given by the derivative
of $p/T^4$ with respect to the temperature.
\begin{eqnarray}
\frac{\Theta^{\mu\mu} (T)}{T^4} \equiv \frac{\epsilon - 3p }{T^4}  =  
T \frac{\partial}{\partial T} (p/T^4) ~~.
\label{delta}
\end{eqnarray}
Since the pressure is given by the logarithm of the partition function, 
$p/T = V^{-1} \ln Z$,
the calculation of the trace anomaly requires the evaluation of straightforward
expectation values.

Using Eq.~(\ref{delta}), the pressure is obtained by integrating
$\Theta^{\mu\mu}/T^5$ over the temperature,
\begin{equation}
\frac{p(T)}{T^4} - \frac{p(T_0)}{T_0^4} = \int_{T_0}^{T} {\rm d}T'
\frac{1}{T'^5} \Theta^{\mu\mu} (T') \;\; .
\label{pres}
\end{equation}
Here $T_0$ is an arbitrary temperature value that is usually chosen in
the low temperature regime where the pressure and other thermodynamical
quantities are suppressed exponentially by Boltzmann factors 
associated with the lightest hadronic states, {\it i.e.}, the pions. We find 
it expedient to extrapolate to $T_0\equiv 0$, in which limit $p/T_0^4\equiv 0$.
Energy ($\epsilon$) and entropy ($s=\epsilon+p$)  densities are then obtained 
by combining results for $p/T^4$ and $(\epsilon -3p)/T^4$.

To calculate basic thermodynamic quantities such as energy density $\epsilon$,
pressure $p$, and the trace anomaly, one needs to know several
lattice $\beta$-functions along the LCP at $T=0$ on which
our calculations have been performed. We determine these $\beta$-functions 
using the same parametrizations for the LCP as in the analysis of the EoS
on lattices with temporal extent $N_\tau = 6$ \cite{milc_eos,rbcBIeos}.
Also, the determination of these $\beta$-functions in the nonperturbative 
regime 
is carried out on the same set of zero temperature lattices used to set 
the temperature scale. Further details are
given in Appendix A.

\begin{figure}[t]
\begin{center}
\epsfig{file=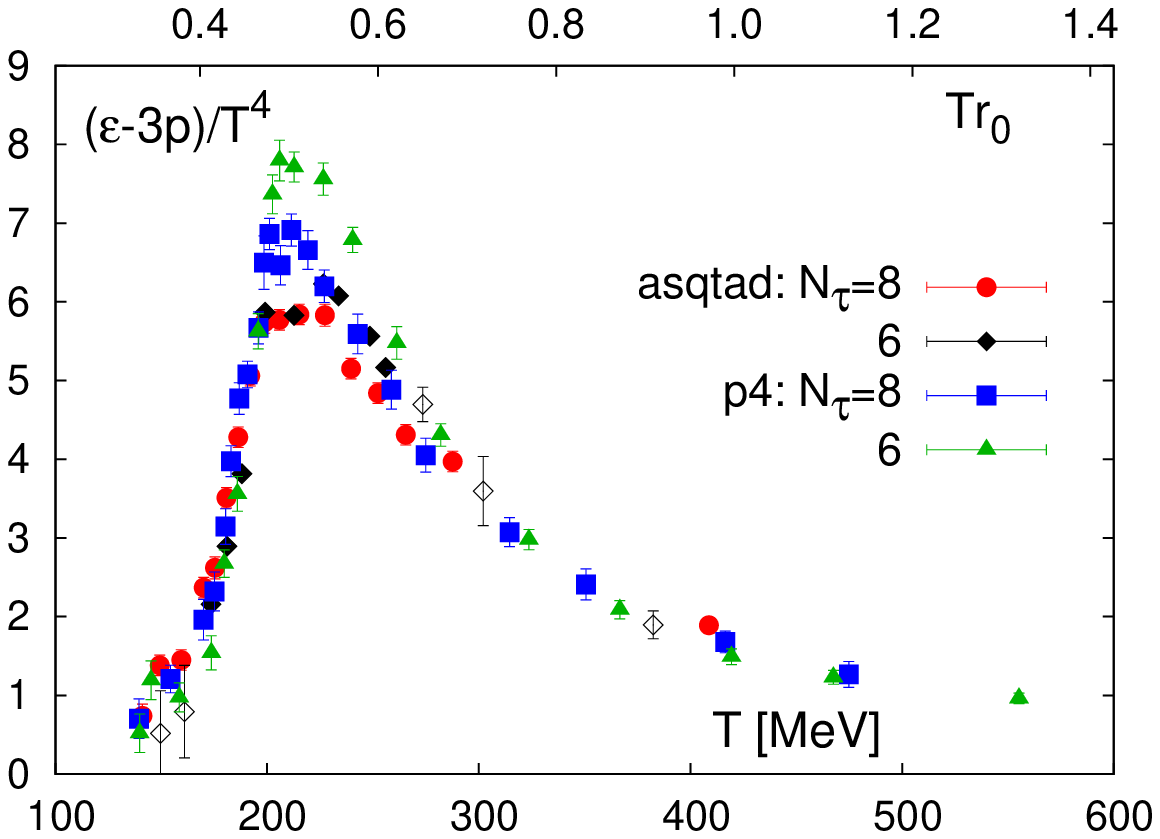,width=10.0cm}\hspace*{-1.4cm}\epsfig{file=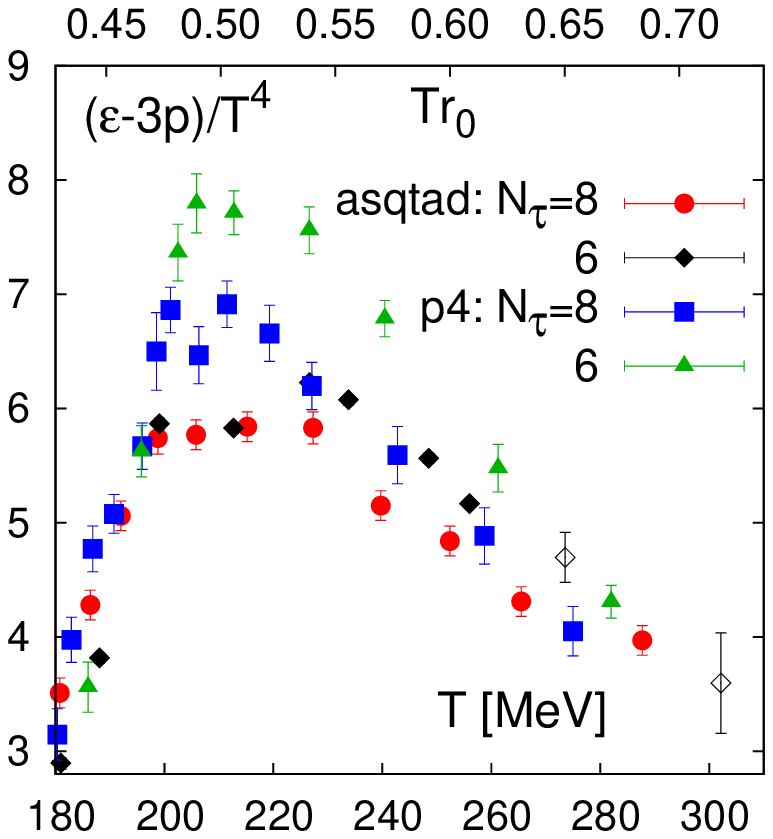,
width=10.0cm}
\end{center}
\caption{\label{fig:e3p}(color online) The trace anomaly $(\epsilon -3p)/T^4$
calculated on lattices with temporal extent $N_\tau =6,~8$. The upper
{\it x}-axis shows the temperature scale in units of the scale parameter
$r_0$ which has been determined in studies of the static quark potential.
The lower {\it x}-axis gives the temperature in units of MeV, which has been
obtained using for $r_0$ the value determined from the level splitting
of bottomonium states, $r_0 = 0.469$~fm \cite{gray}. The right hand figure
shows the region around the maximum of  $(\epsilon -3p)/T^4$, which also is
the temperature region where results obtained with the two different 
discretization schemes show the largest differences.  Open symbols for the 
$N_\tau=6$, asqtad data set denote data obtained with the R~algorithm. All
other data have been obtained with an RHMC algorithm.
}
\end{figure}

In Fig.~\ref{fig:e3p}, we show results for $\Theta^{\mu\mu}/T^4$
obtained with both the asqtad and p4 actions. The new $N_\tau=8$ results
have been obtained on lattices of size $32^3 \times 8$ and the additional
zero temperature calculations, needed to carry out the necessary vacuum
subtractions, have been performed on $32^4$ lattices. The $N_\tau=6$ results 
for the p4 action shown for comparison are taken from \cite{rbcBIeos}. For the asqtad
action, new $N_\tau=6$ RHMC results obtained on $32^3 \times 6$ lattices are shown 
using full symbols while earlier
results obtained on $12^3 \times 6$ lattices with the R~algorithm
\cite{milc_eos} are shown using open symbols.

We find that the results with asqtad and p4 formulations are
in good agreement. In particular, both actions yield consistent
results in the low temperature range, in which $\Theta^{\mu\mu}/T^4$ 
rises rapidly, and at high temperature, $T \gsim 300$~MeV. This is also the case for
the cutoff dependence in these two regimes. At intermediate temperatures,
$200\;{\rm MeV}\lsim T\lsim 300\;{\rm MeV}$, the two actions show 
differences (Fig.~\ref{fig:e3p}(right)).
The maximum in $\Theta^{\mu\mu}/T^4$ is shallower for the asqtad action and shows
a smaller cutoff dependence than results obtained with the p4 action. 

As $(\epsilon -3p)/T^4$ is the basic input for the calculation of all other
bulk thermodynamic observables, we discuss its structure in more detail in
the following subsections.  

\subsubsection{The crossover region}

\begin{figure}[t]
\begin{center}
\epsfig{file=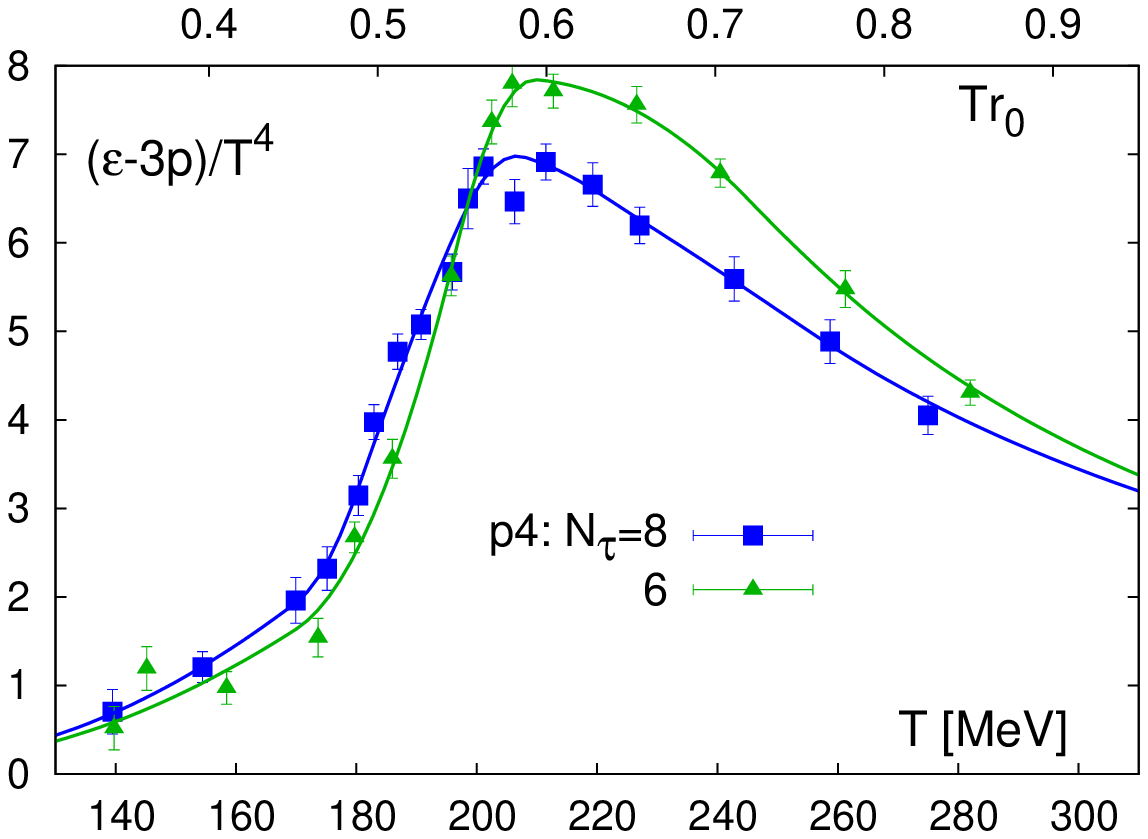,width=7.9cm}
\epsfig{file=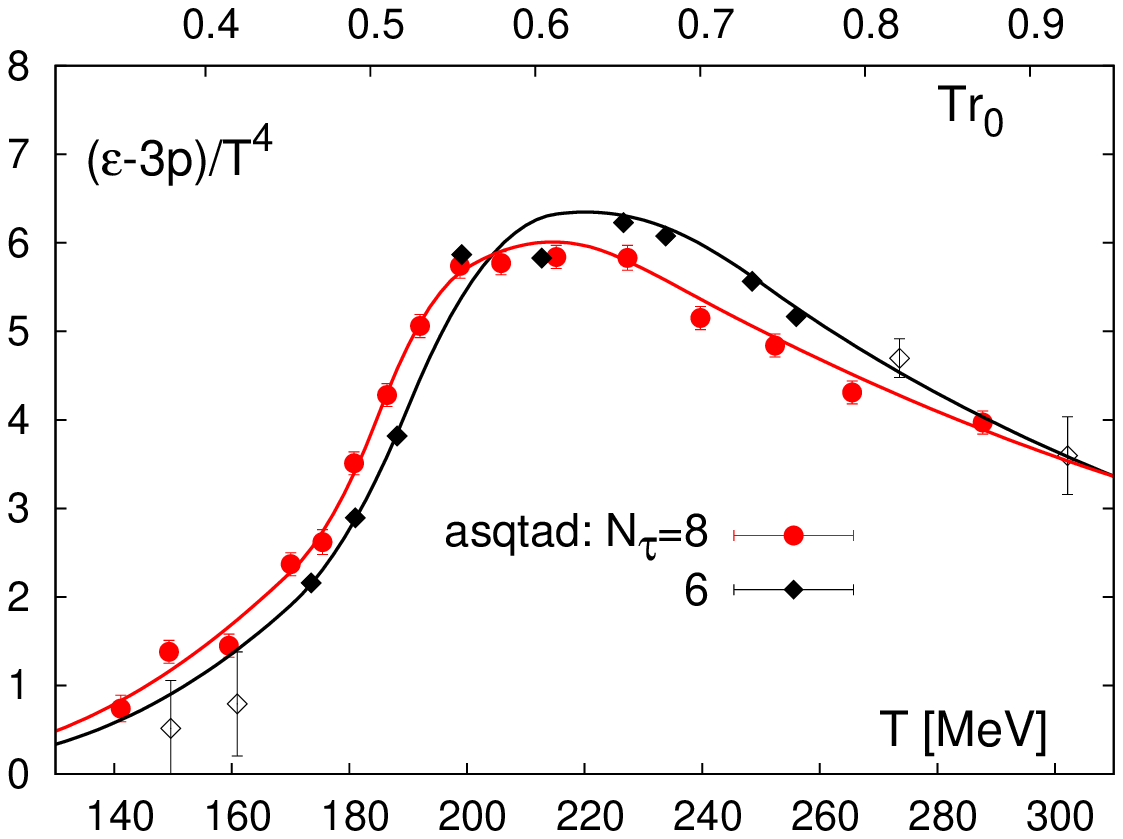,width=7.9cm}
\end{center}
\caption{\label{fig:low} 
(color online) The trace anomaly calculated with the p4 (left) and asqtad (right) actions. 
Shown is a comparison of results obtained on lattices with
temporal extent $N_\tau=6$ and $8$. The curves show interpolations discussed
in the text. Open symbols for the
$N_\tau=6$, asqtad data set denote data obtained with the R~algorithm. All
other data have been obtained with an RHMC algorithm.
}
\end{figure}

In Fig.~\ref{fig:low}, we show results for the trace anomaly for
both actions at $T < 300$~MeV (about $1.5$ times the transition 
temperature\footnote{We stress, however, that the
entire discussion of thermodynamics we present here is completely
independent of any determination of a `transition temperature'.  The
temperature scale is completely fixed through determinations of the 
lattice scale performed at zero temperature.})
and for $N_\tau=6$ and 8 lattices.
In the region $T\le 170$~MeV, the curves shown are exponential fits. 
Above $T=170$~MeV, we divide the data into several intervals and perform 
quadratic interpolations. In each interval, these
quadratic fits have been adjusted to match the value and slope
at the boundary with the previous interval. 
These interpolating curves are then used to calculate the pressure and other 
thermodynamic quantities using Eqs.~(\ref{delta}) and (\ref{pres}). 

The differences between the $N_\tau=6$ and $8$ data in the transition
region can well be accounted for by a shift of the $N_\tau=6$ data
by about $5$~MeV towards smaller temperatures.
This reflects the cutoff dependence of the transition temperature
and may also subsume residual cutoff dependencies of the zero
temperature observables used to determine the temperature scale
in the transition region.
As will become clearer later, we find 
a cutoff dependence of similar magnitude in other observables. 

Such a global shift of scale for the $N_\tau=6$ data set also compensates 
for part of the cutoff dependence seen at higher temperatures. 
It thus is natural to expect that cutoff effects in $(\epsilon -3p)/T^4$
change sign at a temperature close to the peak in this quantity, which occurs
at a temperature $T\gsim 200$~MeV. In the vicinity of this peak, we find the 
largest difference between results obtained with the two actions. The 
cutoff dependence in $(\epsilon -3p)/T^4$ with the p4 action is 
about twice as large as with the asqtad action, and the 
peak height is about 15\% smaller with the
asqtad action than with the p4 action.

\begin{figure}[t]
\begin{center}
\epsfig{file=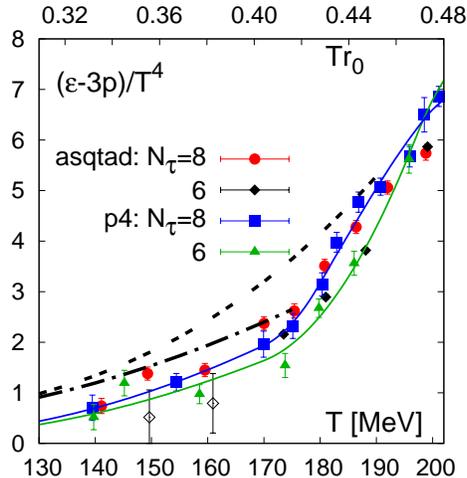,width=9.5cm}
\end{center}
\caption{\label{fig:details} (color online) The trace anomaly at low temperatures
calculated with the asqtad and p4 actions on lattices with temporal
extent $N_\tau=6$ and $8$. 
Open symbols for the
$N_\tau=6$, asqtad data set denote data obtained with the R~algorithm. All
other data have been obtained with an RHMC algorithm.
Solid lines show interpolation 
curves for the p4 action discussed in the text. The dashed and dashed-dotted
curves give the trace 
anomaly calculated in a hadron resonance gas model with two different cuts for the
maximal mass, $m_{max}=1.5$~GeV (dashed-dotted) and $2.5$~GeV (dashed). 
}

\end{figure}

It is of interest to compare how well the thermodynamics of the 
low temperature phase can be characterized by a resonance gas model. In the 
low temperature region, the hadron resonance gas has been observed \cite{pbm}
to give
a good description of bulk thermodynamics. It also is quite successful
in characterizing the thermal conditions met in heavy ion collisions at the 
chemical freeze-out temperature, {\it i.e.}, at the temperature at which hadrons 
again form in the dense medium created in such collisions. 
In Fig.~\ref{fig:details}, we compare the results for $(\epsilon -3p)/T^4$ to 
predictions of the hadron resonance gas model \cite{pbm}, 
\begin{equation}
\left( \frac{\epsilon - 3p}{T^4}\right)_{low-T} =
\sum_{m_i\le m_{max}}
\frac{d_i}{2\pi^2}
\sum_{k=1}^\infty
(-\eta_i)^{k+1}\frac{1}{k}
\left( \frac{m_i}{T}\right)^3 K_1(km_i/T) \; ,
\label{e3plow}
\end{equation}
where different particle species of mass $m_i$ have degeneracy
factors $d_i$ and $\eta_i = -1 (+1)$ for bosons (fermions). The particle
masses have been taken from the particle data book \cite{databook}.
Data in Fig.~\ref{fig:details} show the HRG model results including
resonances up to $m_{max} = 1.5$~GeV (lower curve) and $2.5$~GeV
(upper curve). We find that the $N_\tau=8$ results
are closer to the resonance gas model result and there is a tendency
for the difference between the HRG model and lattice results to 
increase with decreasing temperature. 
This is not too surprising since the light meson sector is not 
well reproduced in current simulations. To quantify to what extent this reflects
deviations from a simple resonance gas model\footnote{Note that the HRG
has to fail in describing higher moments of charge fluctuations in the 
vicinity of the transition temperature \cite{rbcBIfluct}.} 
or can be attributed to the violation of
taste symmetry and/or the still too heavy Goldstone pion
in our current
analysis needs to be analyzed in more detail in the future. Qualitatively, 
the effects of staggered taste symmetry breaking and heavier quarks are to underestimate the energy density 
and pressure.

\subsubsection{High temperature region}
\label{sec:highT}

\begin{figure}[t]
\begin{center}
\epsfig{file=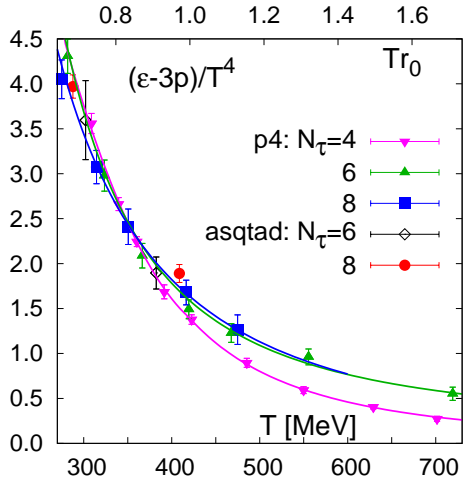,width=9.5cm}
\end{center}
\caption{\label{fig:detailshigh}(color online) The trace anomaly at high temperatures
calculated with the asqtad and p4 actions on lattices with temporal
extent $N_\tau=6$ and $8$. For the p4 action we also show results obtained 
on lattices with temporal extent $N_\tau=4$ \cite{rbcBIeos}. 
Open symbols for the
$N_\tau=6$, asqtad data set denote data obtained with the R~algorithm. All
other data have been obtained with an RHMC algorithm.
Solid curves 
show fits to the data based on Eq.~(\ref{e3phigh}). Fit parameters are given 
in Table~\ref{tab:bc_fit}.
}

\end{figure}

At high temperatures $(\epsilon -3p)/T^4$ will eventually approach zero
in proportion to $g^4(T)\sim 1/\ln^2(T/\Lambda)$ \cite{Boyd}. In the data 
shown in Fig.~\ref{fig:detailshigh} for the temperature range
accessible in our present analysis, {\it i.e.}, $T\; \lsim\; 3.5 T_c$, the 
variation of $(\epsilon -3p)/T^4$ with temperature is, however, 
significantly stronger. 
Following the analysis performed in \cite{rbcBIeos},
we have fit the p4 data at $T > 250$~MeV to the {\it ansatz}
\begin{equation}
\left( \frac{\epsilon - 3p}{T^4}\right)_{high-T} = 
\frac{3}{4}b_0 g^4+\frac{d_2}{T^2} + \frac{d_4}{T^4} \; ,
\label{e3phigh}
\end{equation}
where the first term gives the leading order 
perturbative result and the other terms parametrize nonperturbative
corrections as inverse powers of $T^2$. We find that the $N_\tau=8$ 
data do not extend to high enough temperatures to control the first 
term in the {\it ansatz} given in Eq.~(\ref{e3phigh}). We thus performed fits
to the p4 data sets with $g^2\equiv 0$ and the resulting fit\footnote{We
note that a nonperturbative term proportional to $1/T^4$ arises in
the QCD equation of state from nonvanishing zero temperature condensates.
In fact, our fit results for the coefficient $d_4$ are quite consistent with
commonly used values for the bag parameter. We find
$B^{1/4}=(d_4/4)^{1/4} = (175-225)$~MeV.}
parameters are summarized in Table~\ref{tab:bc_fit}. 
We find that these parameters are 
stable under variation of the fit range and show no 
significant cutoff dependence between $N_\tau=6$ and $8$ data.
The fits for $T\ge 300$~MeV (Table~\ref{tab:bc_fit})
are shown in Fig.~\ref{fig:detailshigh} together with the data
obtained in the high temperature region.
In our calculations with the asqtad action, we did not cover this high 
temperature regime with a sufficient number of data points to perform 
independent fits. However, in Appendix~C we use a modified version 
of Eq.~\ref{e3phigh} to parametrize the equation of state for both p4 
and asqtad for use in hydrodynamic codes.
It is evident from Fig.~\ref{fig:detailshigh} that results obtained with the
asqtad action are in good agreement with the
p4 results.

\begin{table}[t]
\begin{center}
\vspace{0.3cm}
\begin{tabular}{|c|c|c|c||c|c|c|}
\hline
$N_\tau$ & $d_2$ [GeV$^2$] & $d_4$ [GeV$^4$] & $\chi^2$/dof &
$d_2$ [GeV$^2$] & $d_4$ [GeV$^4$] & $\chi^2$/dof\\
\hline
~ &  \multicolumn{3}{|c||}{ $T\ge 300$~MeV} &  
\multicolumn{3}{|c|}{ $T\ge 250$~MeV}\\
\hline
4 &  0.101(6) &  0.024(1) & 1.23 &  0.137(15) &  0.018(2) & 7.38\\
6 &  0.26(5) & 0.005(3) & 1.16 &  0.23(2) & 0.0086(16) & 1.32 \\
8 &  0.22(3) & 0.008(4) & 0.81 &   0.24(2) & 0.0054(17) & 0.66 \\
\hline
\end{tabular}
\end{center}
\caption{Parameters of fits  
to $(\epsilon - 3p)/T^4$ in the region $T\ge 300$~MeV and 
$T\ge 250$~MeV
using the {\it ansatz} given in Eq.~(\ref{e3phigh}) with $g^2\equiv 0$. 
In addition to fit results
for $N_\tau =8$, we also reanalyzed the $N_\tau=6$ data of \cite{rbcBIeos}
setting the constant term in the {\it ansatz} to zero ($g^2\equiv 0$) and 
display again the fit result for the $N_\tau=4$ data given also
in \cite{rbcBIeos}. The 4th and 7th columns give $\chi^2$ per degree of 
freedom in the respective fit intervals.
}
\label{tab:bc_fit}
\end{table}

\subsection{Cutoff dependence of Gluon  and Quark Condensates and continuum 
extrapolation of \boldmath $\Theta^{\mu\mu}$}

In this section, we want to discuss in more detail the various gluonic and 
fermionic contributions to the trace anomaly and use them to analyze the
difference found in calculations performed with the asqtad and p4 actions.
For nonvanishing quark masses, the trace anomaly receives 
contributions $\Theta_F^{\mu \mu}$
that are proportional to the quark mass and contain the quark
condensates, and contributions $\Theta_G^{\mu \mu}$ that do not
vanish in the chiral limit, {\it i.e.}, from the gluon condensate,
\begin{equation}
\Theta^{\mu\mu}/T^4 = \Theta^{\mu\mu}_G/T^4 + \Theta^{\mu\mu}_F/T^4 \; ,
\label{thetaGF}
\end{equation}
with $\Theta^{\mu\mu}_{G,F}$ defined in Eqs.~(\ref{e3pfermion}) and 
(\ref{e3pgluon}) 
in Appendix A (for more details see \cite{rbcBIeos}). 

\begin{figure}[t]
\begin{center}
\hspace*{-1.8cm}\epsfig{file=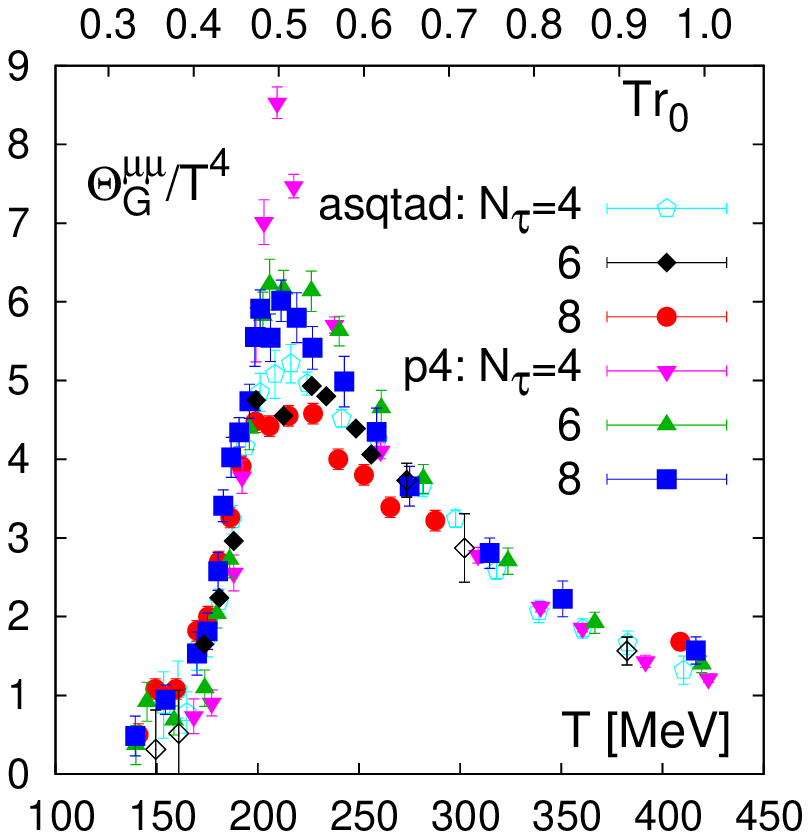,width=12.2cm}\hspace*{-3.7cm}
\epsfig{file=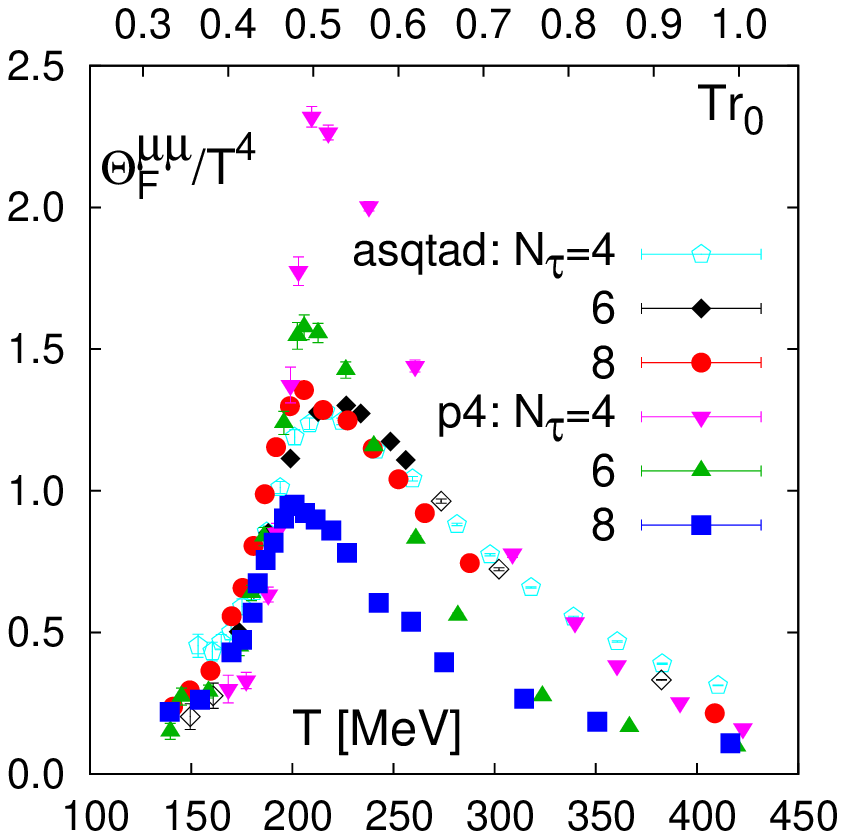,width=12.2cm}
\end{center}
\caption{\label{fig:condensates} (color online) Gluon condensate and quark condensate
contributions to the trace anomaly. Shown are results for the asqtad and p4
actions obtained on lattices of size $N_\tau =4,~6$ and $8$.
Some results shown for the asqtad
and p4 actions on lattices of size $N_\tau =4,~6$ have been taken from
earlier calculations \cite{rbcBIeos,milc_eos}. 
Open symbols for the
$N_\tau=4$ and $6$ asqtad data sets denote data obtained with the 
R~algorithm. All other data have been obtained with an RHMC algorithm.
}
\end{figure} 

In Fig.~\ref{fig:condensates}, we show separately the two contributions 
$\Theta^{\mu\mu}_G/T^4$ and $\Theta^{\mu\mu}_F/T^4$. As seen already 
for $\Theta^{\mu\mu}/T^4$ cutoff effects are generally smaller for the
asqtad action than for the p4 action. By comparing our results obtained
on lattices with temporal extent $N_\tau=6$ and $8$,
we estimate the overall cutoff dependence of $\Theta^{\mu \mu}/T^4$
in the vicinity of its maximum to be about 15\% in calculations
with the p4 action and only half that size in calculations with the
asqtad action.

While for the latter action only the gluonic term
$\Theta^{\mu \mu}_G/T^4$ shows some differences between the $N_\tau=6$
and 8 calculations, in the p4 case the cutoff effects mainly arise
from the fermionic contribution $\Theta^{\mu \mu}_F/T^4$.
Although they are large in this quantity, the quark condensates
contribute less than 15\% to the total trace anomaly.
This contribution reduces to about 5\% at $T \simeq 400$ MeV.
We therefore conclude that at all values
of the temperature, the trace anomaly is strongly dominated by the gluon
condensate contribution which, in turn, receives contributions  
from the quark sector through interactions. 

As pointed out in \cite{rbcBIeos}, the cutoff dependence in 
$\Theta^{\mu\mu}_F/T^4$, seen for the p4 action, mainly arises from the 
structure of the nonperturbative
$\beta$-functions that characterize the variation of bare quark masses along
the LCP, {\it i.e.}, $R_m$. This function appears as a multiplicative factor 
in the 
fermion contribution defined in Eq.~(\ref{e3pfermion}) and approaches
unity in the continuum limit.
The influence of nonperturbative contributions to this prefactor 
thus is shifted to smaller temperatures
as the lattice spacing is reduced, {\it i.e.}, $N_\tau$ is increased,
thereby reducing
the cutoff dependence of $\Theta^{\mu\mu}_F/T^4$ at high temperatures.
Comparing results for $N_\tau=6$ and $8$ one finds, indeed, that results
for $\Theta^{\mu\mu}_F/T^4$ are more consistent for $T\gsim 300$~MeV.
This suggest that the cutoff dependence should be drastically reduced in 
calculations on lattices with temporal extent $N_\tau=12$. 
To check this, we have performed calculations on $32^3\times 12$ lattices 
at 3 $\beta$-values.
Our preliminary results indicate that results are indeed in good agreement
with calculations on $N_\tau =8$ lattices performed at the same value 
of the temperature.

\begin{figure}[t]
\begin{center}
\epsfig{file=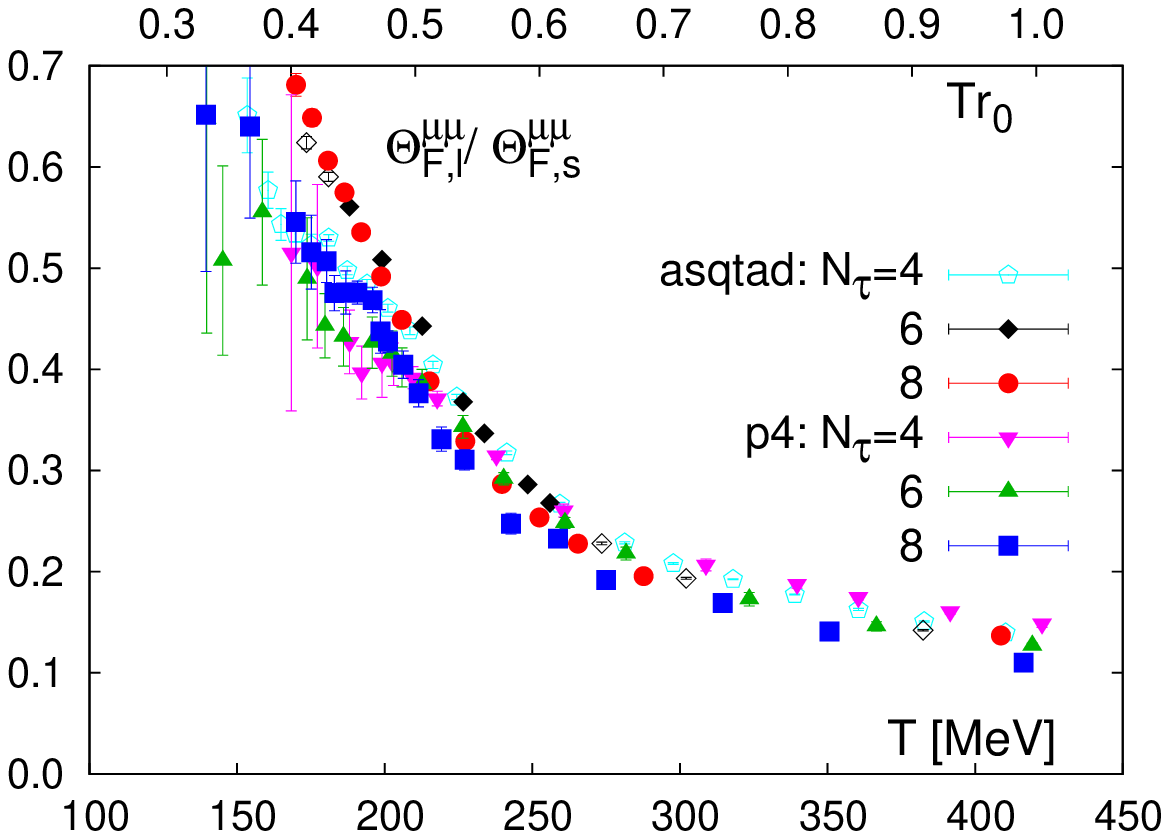,width=12.2cm}
\end{center}
\caption{\label{fig:pbp_l_s} (color online) The ratio of light and strange quark contributions
to the fermionic part of the trace anomaly on lattices with temporal 
extent $N_\tau=4$, 6, and $8$. Some results shown for the asqtad 
and p4 actions on lattices of size $N_\tau =4$, 6 have been taken from 
earlier calculations \cite{rbcBIeos,milc_eos}. 
Open symbols for the
$N_\tau=4$ and $6$ asqtad data sets denote data obtained with the 
R~algorithm. All
other data have been obtained with an RHMC algorithm.
}
\end{figure}

As can be seen in the right hand part of Fig.~\ref{fig:condensates}, results
for $\Theta_F^{\mu\mu}/T^4$ obtained with the asqtad action are systematically 
larger than the 
$N_\tau=8$ results obtained with the p4 action. This is particularly
evident in the high temperature region, $T\gsim 200$~MeV.  
At these values of the temperature 
$\Theta^{\mu\mu}_F/T^4$ is dominated by the contribution of the zero
temperature strange quark condensate and the strange quark mass. The
light quark contribution is suppressed by a factor $m_l/m_s$ and the
thermal contributions are small, relative to the vacuum contributions, due 
to the disappearance of spontaneous symmetry at these temperatures.
The difference in the p4 and asqtad results for $\Theta^{\mu\mu}_F/T^4$
thus can be traced back to the differences in the parametrization of the
LCPs used in calculations with these two different actions. As pointed out
in Sec.~II.A, the strange pseudo-scalar $m_{\bar{s}s}$ is about 15\%
heavier on the asqtad LCP than on the p4 LCP. Through the GMOR relation,
this is related to a larger value of the product 
$m_s\VEV{\bar{\psi}\psi}_{s,0}$. The difference seen in 
Fig.~\ref{fig:condensates}, however, 
drops out in the relative contribution of light
and strange quark condensates to $\Theta^{\mu\mu}_F/T^4$. 
In Fig.~\ref{fig:pbp_l_s} we show the ratio 
\begin{equation}
\frac{\Theta_{F,l}}{\Theta_{F,s}} = 
\frac{2m_l \left( \VEV{\bar{\psi}\psi}_{l,0} -
\VEV{\bar{\psi}\psi}_{l,\tau}\right)}{m_s
\left( \VEV{\bar{\psi}\psi}_{s,0} -\VEV{\bar{\psi}\psi}_{s,\tau}\right) } \; .
\label{ratio_theta_F}
\end{equation}
It is evident from this figure that results for
$\displaystyle{\Theta_{F,l}/\Theta_{F,s}}$ agree quite well in calculations
performed with the asqtad and p4 actions, respectively. This ratio shows much 
less
cutoff dependence than the light and strange quark contributions separately.
This is particularly evident in the case of the p4 action and supports the
observation made before, that the cutoff dependence seen in that case mainly 
arise from the function $R_m$. This prefactor drops out in the ratio 
$\displaystyle{\Theta_{F,l}/\Theta_{F,s}}$. 

As expected, the contribution of the light quark condensates is
suppressed relative to the strange quark contribution because both terms
are explicitly proportional to the light and strange quark masses, respectively.
However, the naive expectation, 
$\displaystyle{\Theta_{F,l}/\Theta_{F,s} \le 2m_l/m_s}$ only holds true
for $T\gsim 300$~MeV, {\it i.e.}, for temperatures larger than 1.5 times the 
transition temperature.
In the transition region, the contribution arising from the light quark
sector reaches about 50\% of the strange quark contribution.

To summarize, we find that a straightforward ${\cal O}(a^2)$ extrapolation 
of the trace anomaly to the continuum limit is not yet appropriate because 
the cutoff dependence arises from different sources which need to be
controlled.
Nonetheless, current $N_\tau=8$ data show that estimates 
for $\Theta^{\mu\mu}/T^4$ in the temperature regime
[$200$~MeV,~$300$~MeV] overestimate the continuum value by not
more than 15\% and less than 5\% for $T>300$~MeV. Furthermore, our 
analysis of the quark contribution to the trace anomaly suggests 
that this contribution is most sensitive to a proper determination
of the LCP that corresponds to physical quark mass values in the 
continuum limit.
Our results suggest that it will be possible to control the cutoff effects 
in the entire 
high temperature regime $T\gsim 200$~MeV through calculations
on lattices with temporal extent $N_\tau =12$.

\section{Thermodynamics: 
Pressure, energy and entropy density, velocity of sound}
\label{sec:thermodynamics}

\begin{figure}[t]
\begin{center}
\hspace*{-0.5cm}\epsfig{file=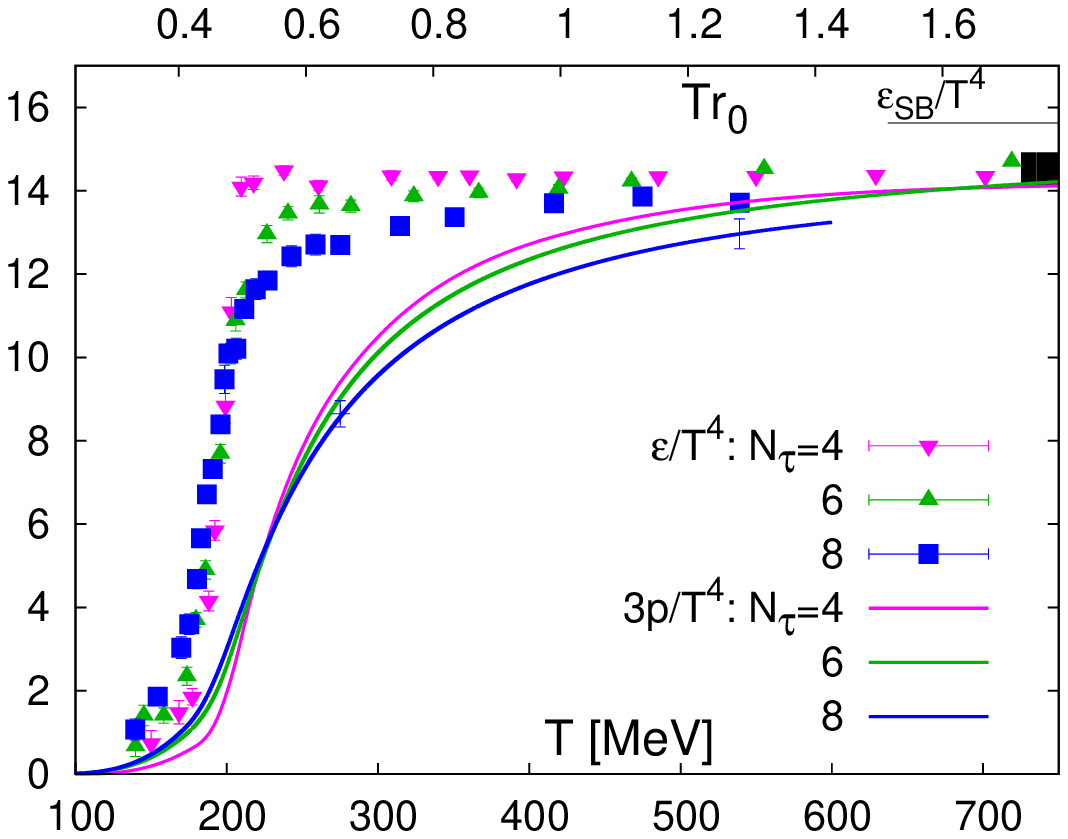,width=9.2cm}\hspace*{-0.5cm}
\epsfig{file=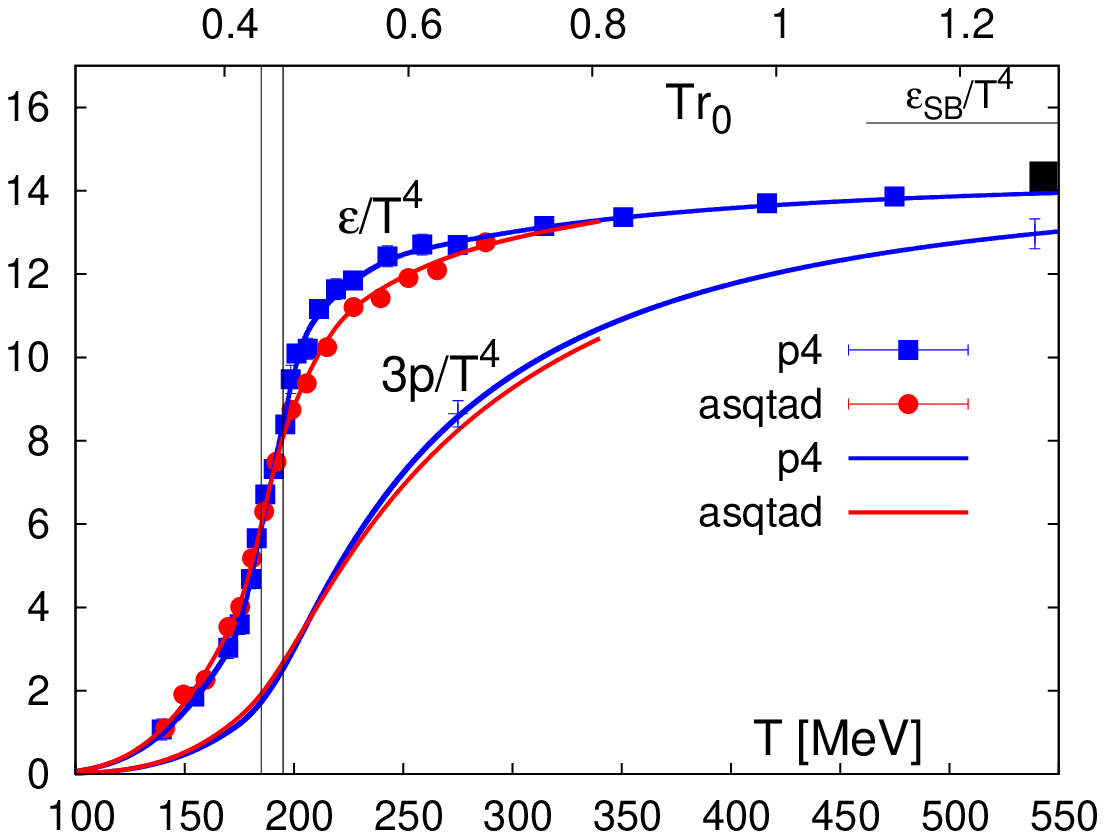,width=9.2cm}
\end{center}
\caption{\label{fig:eos} (color online) Energy density and three times the pressure calculated
on lattices with temporal extent $N_\tau=4,~6$ \cite{rbcBIeos}, and 8
using the p4 action (left). The right hand figure compares results
obtained with the asqtad and p4 actions on the $N_\tau=8$ lattices. 
Crosses with error bars indicate the systematic error on the pressure
that arises from different integration schemes as discussed in the text.
The black bars at high temperatures indicate the systematic shift of data
that would arise from matching to a hadron resonance gas at $T=100$~MeV.
The band indicates the transition region $185\; {\rm MeV} < T < 195\; {\rm MeV}$.
It should be emphasized that these data have not been extrapolated to physical 
pion masses.
}
\end{figure}

We calculate the pressure and energy density from the   
trace anomaly using Eqs.~(\ref{delta}) and (\ref{pres}). 
To obtain the pressure from Eq.~(\ref{pres}), we need to
fix the starting point for the integration. 
In the past, this has been done by choosing a low temperature value   
($T_0 \simeq 100$~MeV) where      
the pressure is assumed to be sufficiently small to be set
equal to zero due to the exponential Boltzmann
suppression of the states. One could also use the hadron resonance gas value for the pressure
at $T_0 =100$~MeV as the starting point for the integration.
The two HRG model fits in Fig.~\ref{fig:details} show that at this temperature the pressure
is insensitive to the exact value of the cutoff $m_{max}$.
We have used both approaches as well as linear interpolations between the
temperatures at which we calculated $\Theta^{\mu\mu}/T^4$ to
estimate systematic errors arising in the calculation of the pressure.
The actual results for $p/T^4$ and other thermodynamic observables shown
in the following have been obtained by starting at $T_0=0$, where we set $p=0$, and
integrating the fits to $\Theta^{\mu\mu}/T^4$ shown in Figs.~\ref{fig:low} and
\ref{fig:details}.
Differences arising from choosing $T_0\simeq 100$~MeV or using linear
interpolations are small and are included in our estimate
of systematic errors.
In Fig.~\ref{fig:eos}, we show our final results for
$\epsilon/T^4$ and $3p/T^4$ obtained in this way. We reemphasize that
$\Theta^{\mu\mu}/T^4$ is what we calculate at a number of temperature values
along the LCP on the lattice, and all other quantities are obtained 
by using fits to this data and then exploiting thermodynamic relations.

\begin{figure}[t]
\begin{center}
\epsfig{file=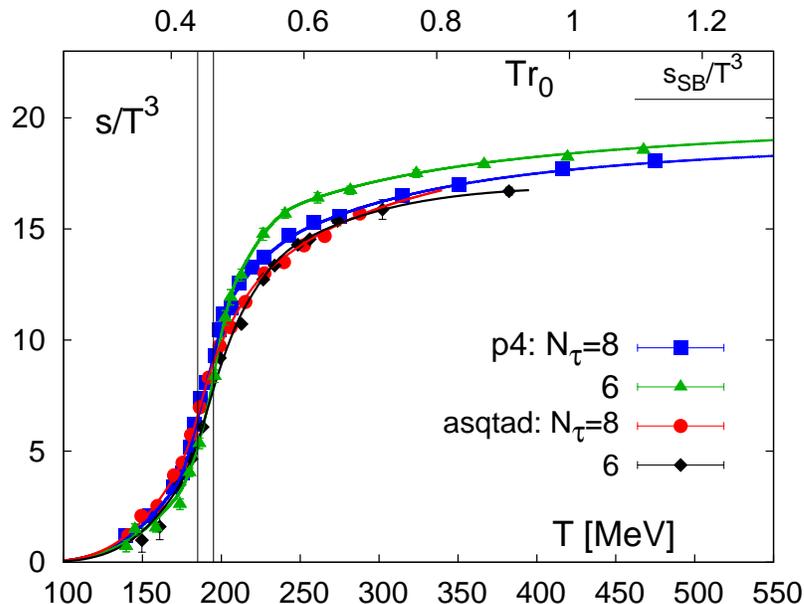,width=12.0cm}
\end{center}
\caption{\label{fig:entropy}(color online) The entropy density obtained on 
lattices with temporal extent $N_\tau =6$ \cite{milc_eos,rbcBIeos} and $8$.
The band indicates the transition region $185\; {\rm MeV} < T < 195\; {\rm MeV}$.
}
\end{figure}

Choosing $T_0=100$~MeV for the starting point of the integration, and        
adding the resonance gas pressure                             
at this temperature to the lattice results gives a global shift of the           
pressure ($3p/T^4$) and energy density ($\epsilon/T^4$) curves by 0.8.
This is indicated by the filled box 
in the upper right hand part of Fig.~\ref{fig:eos}. 
Differences in the results that arise from the different integration schemes
used to calculate the pressure are of similar magnitude. Typical error bars 
indicating the magnitude of this systematic error on $3p/T^4$ are shown in       
the right hand Fig.~\ref{fig:eos} at $T=275$~MeV and $T=540$~MeV. In this
figure, we also compare results obtained                                          
on the $N_\tau=8$ lattices with the p4 and asqtad actions.
The agreement between the two data sets is good in the
entire temperature range that is common. This is a consequence
of the good agreement for the trace
anomaly, from which $\epsilon/T^4$ and $p/T^4$ are derived.
The same discussion applies to the 
entropy density, $s/T^3 = (\epsilon +p)/T^4$, shown in 
Fig.~\ref{fig:entropy} which is obtained by 
combining the results for energy density and pressure.
The comparison of bulk thermodynamic observables ($p,\; \epsilon,\; s$) calculated
on $N_\tau=8$ lattices yields a 
consistent picture for the p4 and asqtad actions. To quantify systematic 
differences, we consider the combination
\begin{equation}
\Delta O \equiv 2\frac{O_1 -O_2}{O_1+O_2} \; ,
\label{relative}
\end{equation}
where $O_1$ ($O_2$) are estimates with the p4 (asqtad) action. We find that 
the relative difference in the pressure $\Delta p$ for temperatures above the crossover region, 
$T\gsim 200$~MeV, is less than 5\%.
This is also the case for energy and entropy density for $T\gsim 230$~MeV with 
the maximal relative difference increasing to 10\% at 
$T\simeq 200$~MeV.  This is a consequence of the difference in the height of the peak 
in $(\epsilon -3p)/T^4$ as shown in Fig.~\ref{fig:e3p}.
Estimates of systematic differences in the low temperature regime are less reliable
as all observables become small rapidly. Nonetheless, the relative differences 
obtained using the interpolating curves shown in Figs.~\ref{fig:eos}
and \ref{fig:entropy} are less than 15\% for $T\gsim 150$~MeV.  
We also find that the cutoff errors between $aT=1/6$ and $1/8$ lattices 
are similar for the p4 action, {\it 
i.e.}, about 15\% at low temperatures and 5\% for $T\gsim 200$~MeV. For 
calculations with the asqtad action, statistically significant cutoff dependence 
is seen only in the difference $(\epsilon -3p)/T^4$.

We conclude that cutoff effects in $p/T^4,\;\epsilon/T^4$ and $s/T^3$ are
under control in the high temperature regime $T\gsim 200$~MeV. 
Estimates of the continuum limit obtained by extrapolating data from 
$N_\tau=6$ and $8$ lattices differ from  the values on $N_\tau=8$ lattices 
by at most 5\%. These results imply that residual ${\cal O}(a^2g^2)$ 
errors are small with both p4 and asqtad actions. 

We note that at high temperatures the results for the pressure presented 
here are  by 20\% to 25\% larger than those reported in \cite{aoki}. 
These latter results have been obtained on lattices with temporal 
extent $N_\tau=4$ and $6$ using the stout-link action. As this action is 
not ${\cal O}(a^2)$ improved, large cutoff effects show up at high temperatures. 
This is well known to happen in the infinite temperature ideal gas limit, 
where the cutoff corrections can be calculated analytically. 
For the stout-link action on the coarse $N_\tau = 4$ and 6 lattices the
lattice Stefan-Boltzmann limits are a factor $1.75$ and $1.51$ higher than the
continuum value. In Ref.~\cite{aoki} it has been attempted to correct for these 
large cutoff effects by dividing the numerical simulation results at finite
temperatures by these factors obtained in the infinite temperature limit.
As is known from studies in pure $SU(N)$ gauge theories \cite{Boyd}, this 
tends to over-estimate the actual cutoff dependence.

\begin{figure}[t]
\begin{center}
\epsfig{file=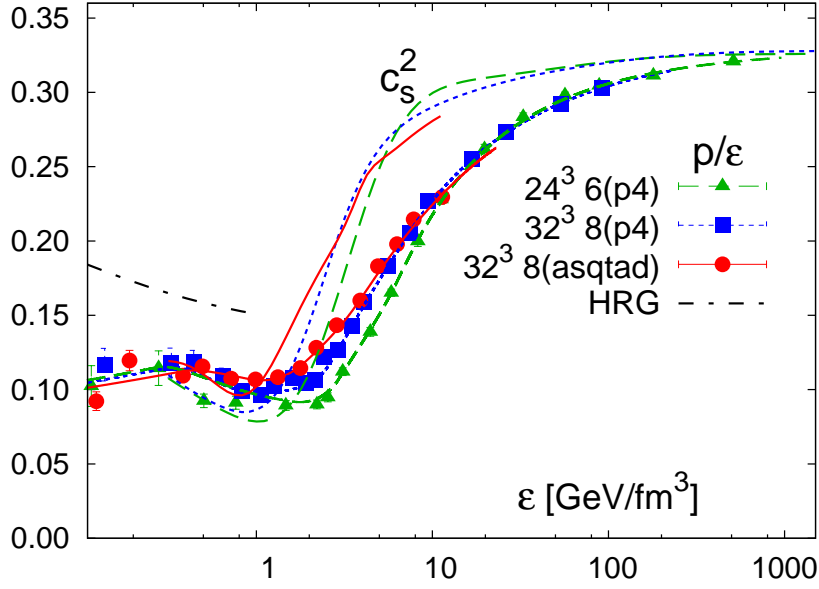,width=12.0cm}
\end{center}
\caption{\label{fig:povere}(color online) Pressure divided by energy density ($p/\epsilon$) 
and the square 
of the velocity of sound ($c_s^2$) calculated on lattices with temporal extent 
$N_\tau =6$ (p4, \cite{rbcBIeos}) and $N_\tau =8$ using the p4 as 
well as the asqtad action.  Lines without data points give the square of the 
velocity of sound calculated analytically from 
Eq.~(\ref{sound}) using the interpolating curves for $\epsilon/T^4$ and $p/T^4$.
The dashed-dotted line at low temperatures gives the result for $p/\epsilon$ 
from a hadron resonance gas (HRG) calculation using $m_{max}=2.5$~GeV.
}
\end{figure}

Finally, we discuss the calculation of the velocity of sound from the basic
bulk thermodynamic observables discussed above. The basic quantity is the
ratio of pressure and energy density $p/\epsilon$ shown in 
Fig.~\ref{fig:povere}, which is obtained from the ratio 
of the interpolating curves for $(\epsilon -3p)/T^4$ and $p/T^4$. 
On comparing results from $N_\tau=6$ and $8$ lattices with the p4 action,
we note that a decrease in the maximal value of $(\epsilon -3p)/T^4$ with $N_\tau$ 
results in a weaker temperature dependence 
of $p/\epsilon$ at the dip (corresponding to the peak in 
the trace anomaly), somewhat larger values  in
the transition region and a slower rise with temperature after the dip. 

From the interpolating curves, it is also straightforward to derive
the velocity of sound,
\begin{equation}
c_s^2 = \frac{{\rm d} p}{{\rm d}\epsilon} = \epsilon 
\frac{{\rm d} (p/\epsilon)}{{\rm d}\epsilon} + \frac{p}{\epsilon}\; .
\label{sound}
\end{equation}
Again, note that the velocity of sound is not an independent
quantity but is fixed by the results for $\Theta^{\mu\mu}/T^4$.
The determination of $c_s^2$ is 
sensitive to the details of the interpolation used to fit the data for $\Theta^{\mu\mu}/T^4$
obtained at a discrete set of temperature values. In fact, it was this sensitivity 
that motivated us to use a smooth curve with a continuous first 
derivative for the interpolation
of $\Theta^{\mu\mu}/T^4$ in the entire temperature range. We consider the 
spread in the curves obtained from 
the p4 and asqtad calculations as indicative of the uncertainty
in the current determination of the velocity of sound from the QCD equation of state
(see also Appendix C).

\section{Deconfinement and chiral symmetry restoration}

In the previous sections, we discussed the thermodynamics of QCD
with almost physical values of the quark masses. Data in Figs.~\ref{fig:eos} and \ref{fig:entropy} show that 
the transition from low to high temperatures occurs in a narrow 
temperature range: $T\in [180\;{\rm MeV},200\;{\rm MeV}]$. This represents 
a crossover and not a true phase transition caused by a singularity
(nonanalyticity)
in the QCD partition function. We expect such a singularity to exist in 
the chiral limit. For QCD with a physical value of the strange quark
mass, it is not yet settled whether 
a true phase transition occurs only at strictly zero light quark masses
($m_l\equiv m_{crit}=0$) or at small but nonzero value 
($m_l\equiv m_{crit}>0$). 
In the latter case, the second order phase transition will belong to the 
universality class of a three dimensional Ising model, while in the former 
case, it is expected to belong to the universality class of three dimensional, 
$O(4)$ symmetric spin models\footnote{For
the discussion of deconfining and chiral symmetry restoring aspects
of the QCD transition in this paper it does not matter whether 
$m_{crit}$ is nonzero or zero.}.  For light quark 
masses $m_l\le m_{crit}$ different observables that are defined through derivatives
of the QCD partition function with respect to either the light quark mass or
a temperature-like variable 
will give unambiguous signals for the occurrence and location of
the phase transition. {For $m_l > m_{crit}$, in the absence of a 
singularity, }
the determination of
a pseudo-critical temperature that characterizes the crossover
may depend on the observable used for its determination. It then
becomes a quantitative question as to what extent different observables
remain sensitive to the singular part of the free energy
density $f_{sing}$ that controls thermodynamics in the vicinity of the 
phase transition temperature at $m_{crit}$. We write the free energy as
\begin{equation}
f = -\frac{T}{V} \log Z\equiv f_{sing}(t,\bar{m})+ f_{reg}(T,m_q) \; ,
\label{free_energy}
\end{equation}
with the reduced mass and temperature variables, 
\begin{equation}
\bar{m}\equiv |m_l-m_{crit}| \;\;{\rm and}\;\;  
t = \left|\frac{T-T_c}{T_c}\right| + c\left(\frac{\mu_l}{T_c}\right)^2  \; .
\label{reduced} 
\end{equation}
Note that in the definition of the reduced temperature $t$,  
its dependence on the light quark chemical potential $\mu_l$
in the vicinity of the critical point $(t,\bar{m})\equiv (0,0)$
was taken into account\footnote{We suppress here a possible but small coupling
to the strange quark chemical potential.}.
To leading order, the reduced temperature depends quadratically on 
$\mu_l$, while it is linear in the temperature itself.

Derivatives of the free energy with respect to quark masses define the
light and strange quark chiral condensates,
\begin{equation}
\langle \bar{\psi}\psi \rangle_q = \frac{T}{V}
\frac{\partial \ln Z}{\partial m_q} \; , \;\; q=l,\; s,
\label{chiral}
\end{equation}
while derivatives with respect to temperature
give the bulk thermodynamic quantities discussed in the previous sections.
Here $m_l$ refers to one of the degenerate light up or down quark 
masses  and the condensates defined in Eq.~(\ref{chiral}) are
one-flavor condensates. The derivative of the chiral condensate with 
respect to the quark mass defines the chiral susceptibilities
$\chi_{m,q} \sim \partial^2\ln Z / \partial m_{q}^2$. The divergence of
$\chi_{m,q}$ at $T_c$ in the chiral limit is an unambiguous signal
of the chiral phase transition. 
In addition, the fluctuations of Goldstone modes also induce divergences in the
chiral limit  for $T\le T_c$
\cite{goldstone}. Thus $\chi_{m,q}$ in the chiral limit is finite only
for $T>T_c$.

In the vicinity of $T_c$, where thermodynamics is dominated
by the singular part of the partition function, $n$ derivatives with respect 
to temperature $T$ are equivalent to $2n$ derivatives with respect
to the light quark chemical potential. Second derivatives with respect to light 
and strange quark chemical potentials define quark number
susceptibilities,
\begin{equation}
\frac{\chi_{q}}{T^2} = \frac{1}{VT^3} 
\frac{\partial^2\ln Z}{\partial(\mu_{q}/T)^2} \; ,\;\; q=l,\; s \ .
\label{chi_q}
\end{equation}
Also, here $q=l$ refers to either the light
up or down quark chemical potential, {\it i.e.}, $\chi_l$ defines the
fluctuations of a single light quark flavor. One therefore expects 
the quark
number susceptibilities to exhibit a temperature dependence similar to 
that found for the energy density and  the fourth order cumulant
$c_4 \sim \partial^4\ln Z / \partial(\mu_{q}/T)^4$ to behave
like a specific heat $c_V$~\cite{rbcBIfluct}. Also, the position of the peak in the 
fourth order cumulant is sensitive to the
crossover region seen, for example, in the temperature dependence of $\epsilon /T^4$.

At $\bar{m}=0$, the temperature
dependence of all the observables discussed above is sensitive to the location 
of $T_c$. Near $T_c$, their temperature dependence
reflects the singular structure of the partition function, 
\begin{eqnarray}
\langle \bar{\psi}\psi \rangle_l \sim t^{\beta} \; ({\rm for}\; T\le T_c)
&\;\; ,\;\; & 
\frac{\chi_{l}}{T^2} \sim \frac{\epsilon}{T^4} \sim  A_\pm t^{1-\alpha} +
regular
\label{deriv1}\\
\chi_{m,l} \sim t^{-\gamma} \; ({\rm for}\; T\ge T_c) 
&\;\; , \;\;& 
c_4 \sim c_V \sim A_\pm t^{-\alpha} + regular 
\label{scaling}
\end{eqnarray}
where $\alpha,\; \beta,\; \gamma$ are critical exponents of the relevant 
universality class and $A_\pm$ are proportionality constants that may differ
below and above $T_c$.  
The specific heat, and thus $c_4$, diverge at a generic
second order phase transition, for example those in the Ising 
universality class, whereas these quantities only develop a pronounced peak in
$O(N)$ symmetric models for which the critical exponent $\alpha$ exhibits 
unconventional behavior and is negative.

In the following subsections, we will focus on the temperature
dependence of the chiral condensate and the quark number susceptibilities
(Eq.~(\ref{deriv1})),
which in the chiral limit probe the chiral symmetry restoring and deconfining 
features of the QCD phase transition. In particular, we will discuss the extent to which the
temperature dependence of these quantities remains correlated away from $\bar m = 0$.

\subsection{Deconfinement}

The bulk thermodynamic observables $p/T^4$, $\epsilon/T^4$, and $s/T^3$ discussed
in the previous sections are sensitive to the change from hadronic to 
quark-gluon degrees of freedom that occur during the QCD transition; 
they thus reflect the deconfining features of this transition. 
The rapid increase 
of the energy density, for instance, reflects the liberation of light
quark degrees of freedom; the energy density increases from values close
to that of a pion gas to almost the  value of an ideal gas of  massless
quarks and gluons.
In a similar vein, the temperature dependence of quark number susceptibilities 
gives information on 
thermal fluctuations of the degrees of freedom that carry
a net number of light or strange quarks, {\it i.e.},
$\chi_q \sim \langle N_q^2\rangle$, with $N_q$ denoting the net number
of quarks carrying the charge $q$. Quark number susceptibilities change
rapidly in the transition region as the carriers of charge, strangeness
or baryon number are heavy hadrons at low temperatures but much lighter  
quarks at high temperatures. 
In the continuum and infinite temperature limit, these susceptibilities 
approach the value for an ideal massless one flavor quark 
gas,
{\it i.e.}, $\lim_{T\rightarrow\infty} \chi_q/T^2 = 1$.
At low temperatures, however,
they reflect the fluctuations of hadrons carrying net light quark (up or down)
or strangeness quantum numbers. In the zero temperature limit,
$\chi_s/T^2$ receives contributions only from the lightest 
hadronic state that carries strangeness, $\chi_s/T^2 \sim \exp (-m_K/T)$,
while $\chi_l/T^2$ is sensitive to pions, $\chi_l/T^2 \sim \exp (-m_\pi/T)$.
Note that $\chi_l/T^2$ is directly sensitive to the singular structure of
the QCD partition function in the chiral limit.

\begin{figure}
\epsfig{file=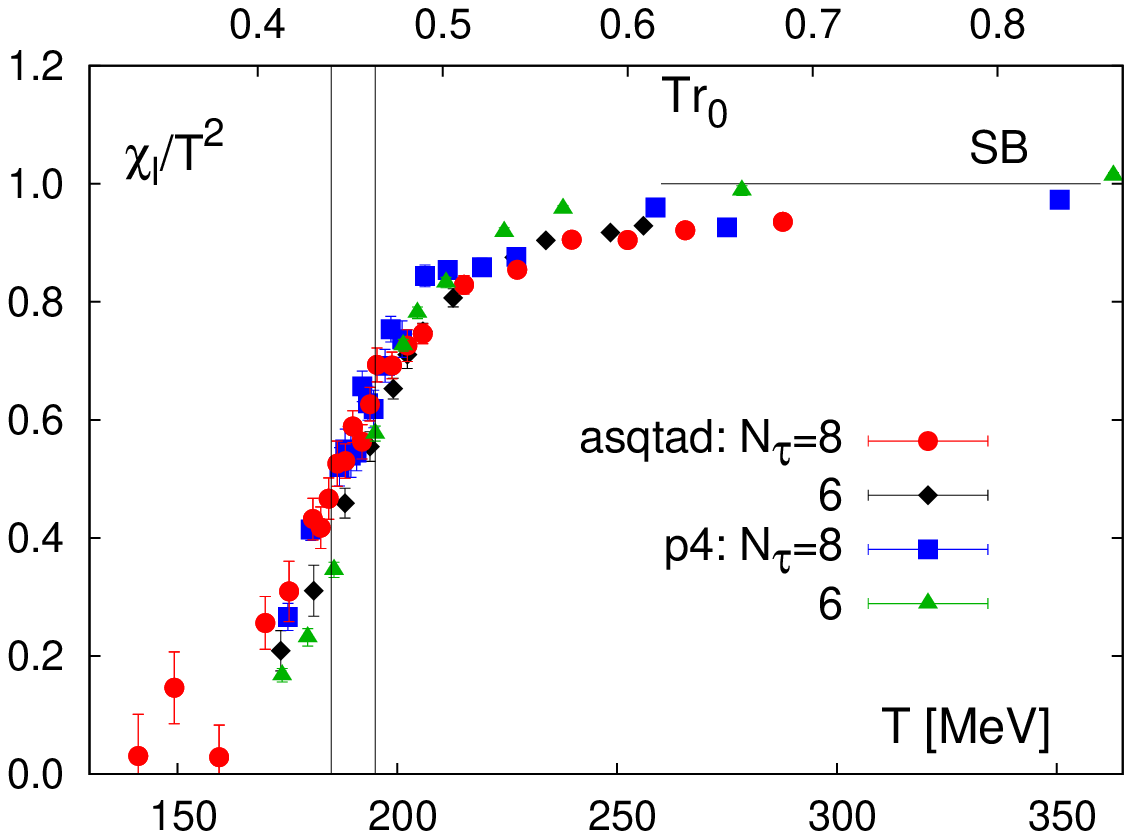,width=8.0cm}
\epsfig{file=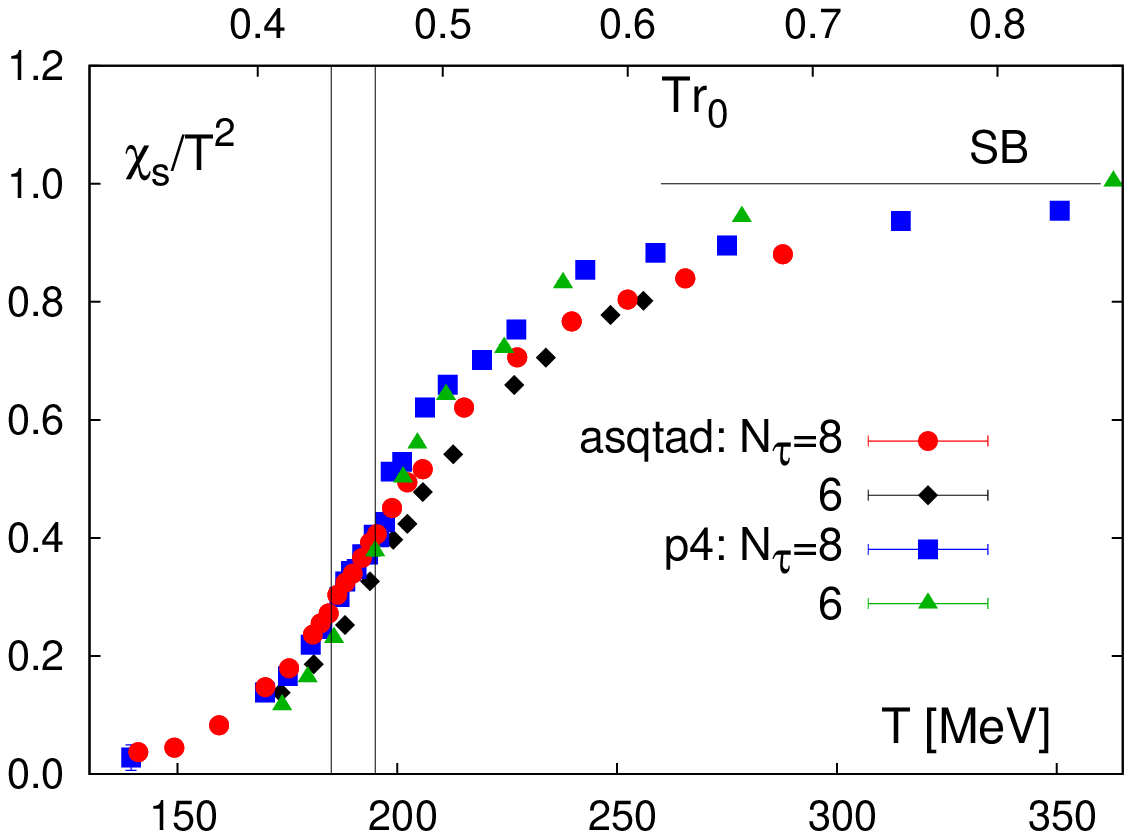,width=8.0cm}
\caption{(color online) The light (left) and strange (right) quark number susceptibilities 
calculated on lattices with temporal extent $N_\tau =6$ 
and $8$. The $N_\tau =6$ results for the p4 action are taken 
from \cite{rbcBIeos}.
The band corresponds to a temperature 
interval $185\;{\rm MeV} \le T \le 195\;{\rm MeV}$. 
}
\label{fig:qno}
\end{figure}

In Fig.~\ref{fig:qno}, we show results for the temperature dependence of the
light and strange quark number susceptibilities. For the ${\cal O}(a^2)$ improved
p4 and asqtad actions, deviations from the continuum result
are already small in calculations on lattices with temporal extent $N_\tau =8$.
The continuum ideal gas value, $\chi_q/T^2 = 1$, is thus a good guide for 
the expected behavior of $\chi_l/T^2$ in the high temperature limit. As can be 
seen from the figure, cutoff effects are indeed small at high temperature. 
In fact, we observe that above the transition, in particular for 
temperatures up to about 1.5 times the transition temperature, differences 
between results obtained with the asqtad and p4 actions are larger than the 
cutoff effects seen 
for each of these actions separately. This is similar to what has been found 
in calculations of the energy density discussed in the previous section. 

\begin{figure}
\epsfig{file=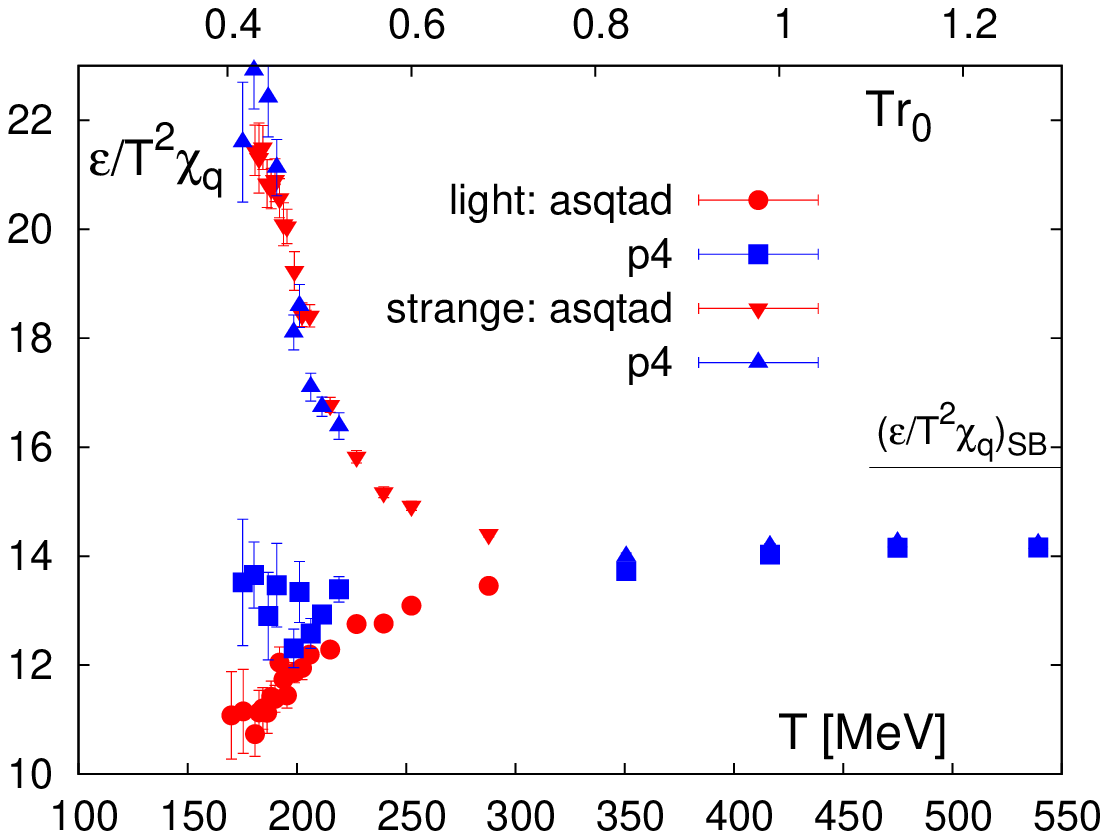,width=8.0cm}
\epsfig{file=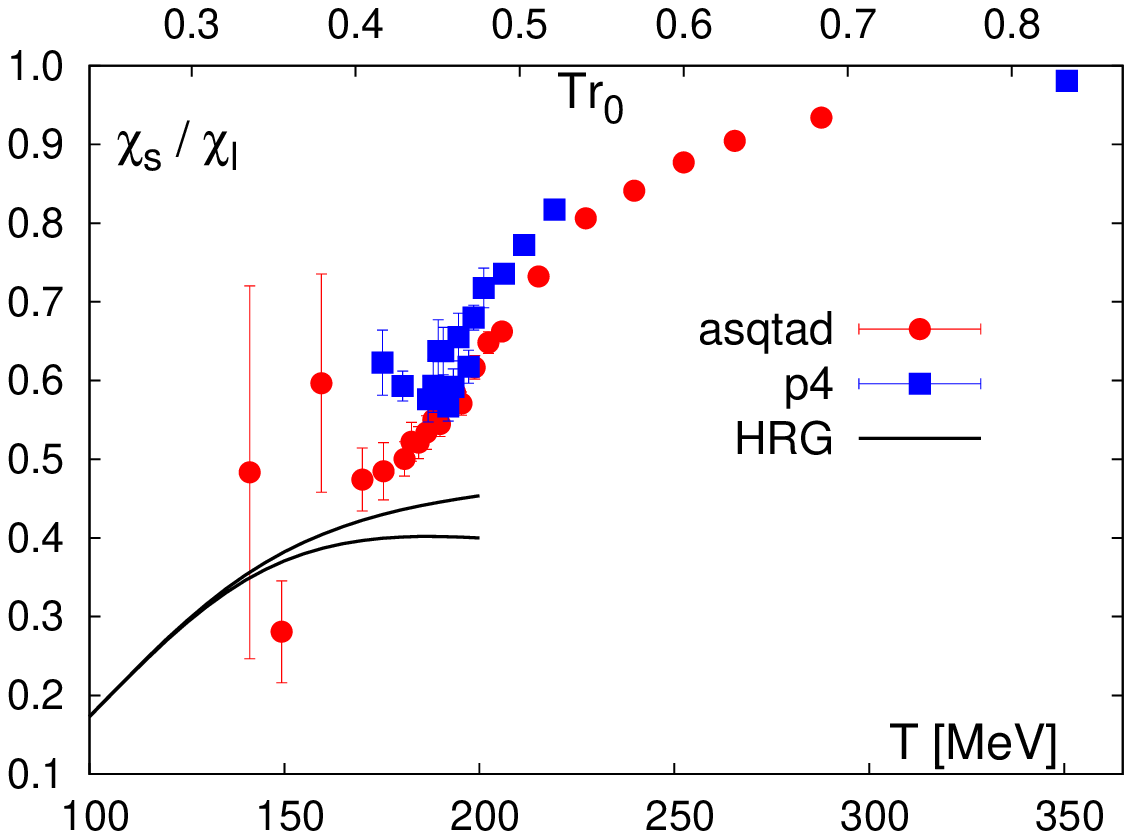,width=8.0cm}
\caption{(color online) The ratio of energy density and quark number susceptibilities (left)
and the ratio of strange and light quark number susceptibilities (right)
calculated on lattices with temporal extent $N_\tau =8$.
For the energy densities the interpolating curves shown in 
Fig.~\protect\ref{fig:eos} have been used.
Curves in the right hand figure show results for a hadron resonance gas including
resonance up to $m_{max}=1.5$~GeV (upper branch) and $2.5$~GeV (lower  branch),
respectively. 
}
\label{fig:ratio}
\end{figure}

The rise of $\chi_l/T^2$ in the transition region is compatible with the
rapid rise of the energy density. The strange quark number susceptibility, on the
other hand, rises more slowly in the transition region. The astonishingly strong
correlation between energy density and light quark number susceptibility is 
evident from the ratio $\epsilon/(T^2\chi_l)$ shown in Fig.~\ref{fig:ratio}(left).
The ratio  $\epsilon/(T^2 \chi_l)$ varies in the transition region by about 15\%.
This is in contrast to the ratio $\epsilon/(T^2 \chi_s)$ which starts to increase
with decreasing temperature already
at $T\simeq 250$~MeV and rises by about 50\% in the transition region. In fact,
unlike  $\epsilon/(T^2\chi_l)$, the ratio  $\epsilon/(T^2\chi_s)$ will diverge in the 
zero temperature limit as $\chi_s$ goes to zero faster than $\epsilon$ 
since it is not sensitive
to the light quark degrees of freedom that contribute to the energy density.
This lack of sensitivity of $\chi_s/T^2$ to the singular structure 
of the QCD partition function suggests that $\chi_s/T^2$ is not a good quantity to use to 
define the pseudo-critical temperature.

The difference in light and strange quark masses 
plays an important role in the overall magnitude of quark number
fluctuations up to almost twice the transition temperature.
This is seen in  Fig.~\ref{fig:ratio}(right), where we show the ratio 
$\chi_s/\chi_l$. 
At $T\gsim 300$~MeV the ratio $\chi_s/\chi_l$ comes close to the 
infinite temperature ideal gas value  $\chi_s/\chi_l=1$. Deviations from
this may be understood as a thermal effect that arises even 
in a noninteracting gas from just the differences in quark masses. 
However, as has been
discussed in \cite{rbcBIfluct}, this clearly is not possible in the transition 
region, $200~{\rm MeV}\lsim T \lsim 300~{\rm MeV}$ where, upon cooling, 
strangeness 
fluctuations decrease strongly relative to light quark fluctuations
as $T$ decreases. 
At the transition 
temperature, $\chi_s/\chi_l$ is only about $1/2$ and has the tendency
to go over smoothly into values extracted from the HRG model. In the
low temperature hadronic region, the ratio $\chi_s/\chi_l$ drops 
exponentially as strangeness fluctuations are predominantly carried
by heavy kaons whereas  the light quark fluctuations are carried by
light pions.

\subsection{Chiral symmetry restoration}

In the limit of vanishing quark masses, the chiral condensate
$\langle \bar{\psi}\psi \rangle_l$ introduced in 
Eq.~(\ref{chiral}) is an order parameter for spontaneous 
symmetry breaking; it stays
nonzero  at low temperature and vanishes above a critical temperature $T_c$.
Chiral symmetry is broken spontaneously for $T<T_c$. 

\begin{figure}
\epsfig{file=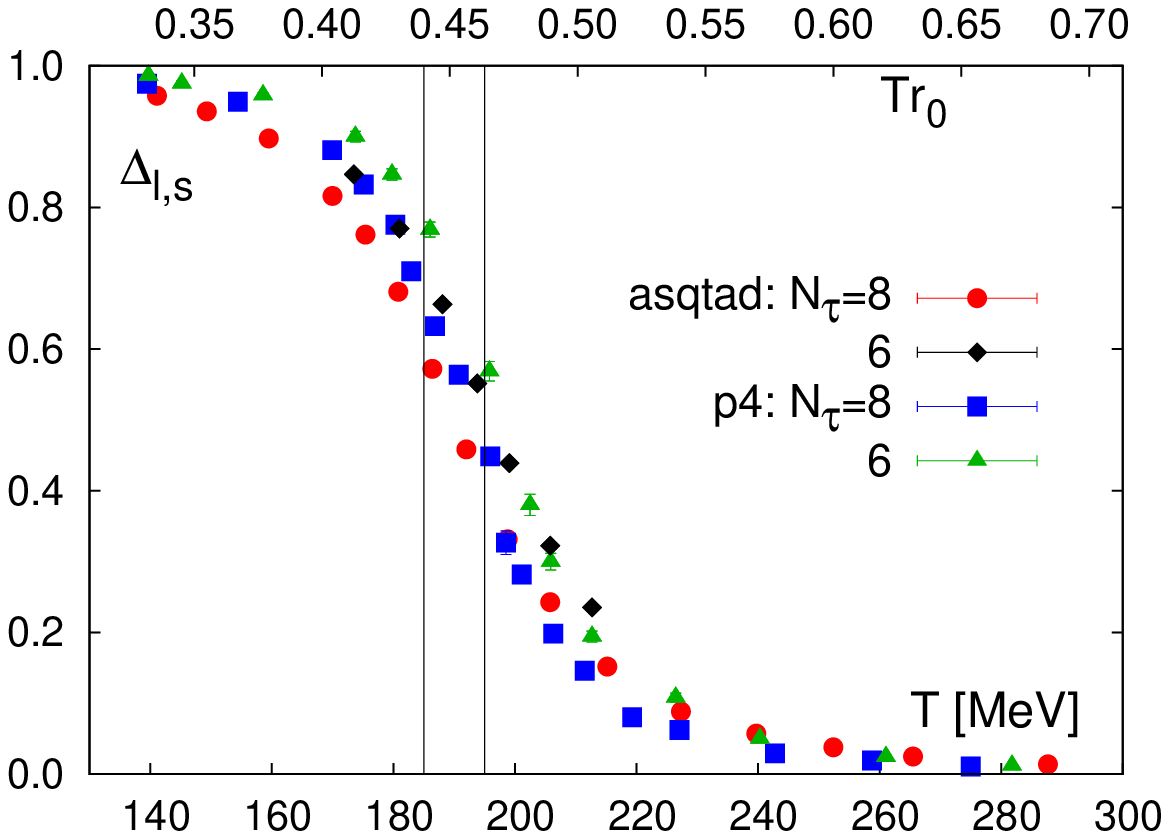,width=8.0cm}
\epsfig{file=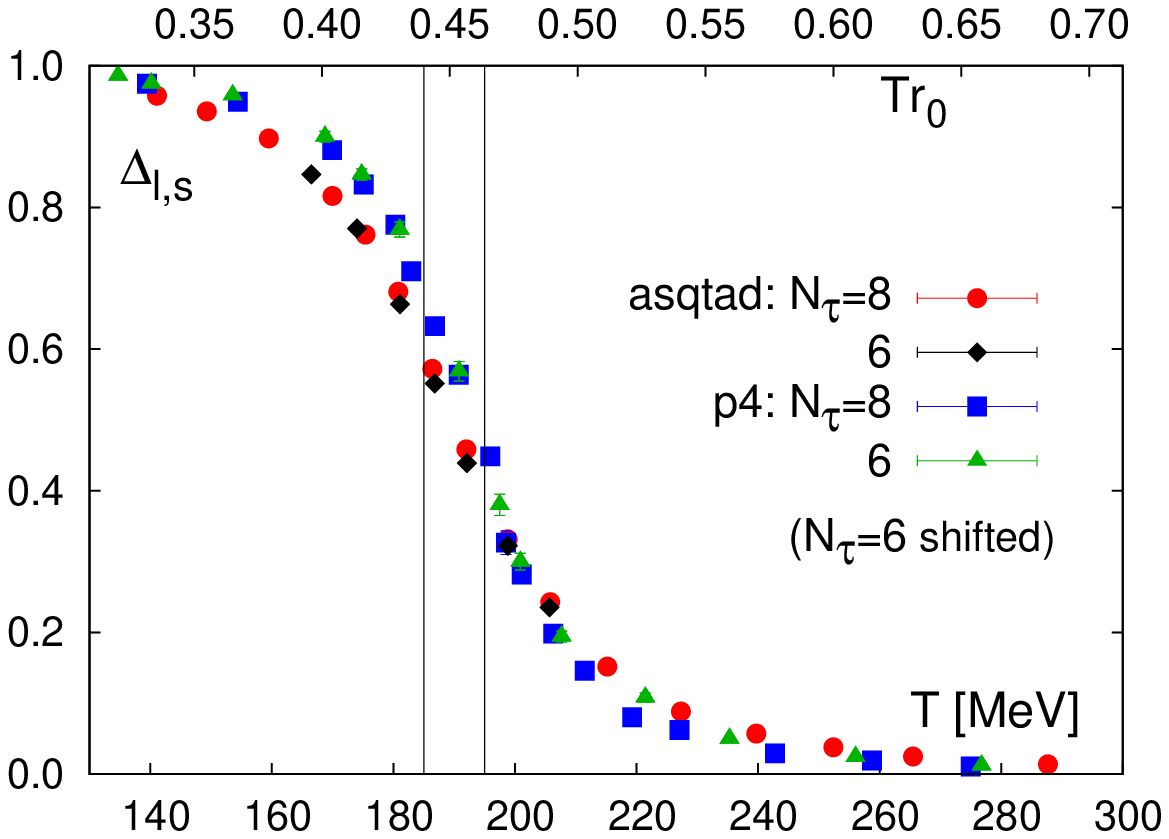,width=8.0cm}
\caption{(color online) The subtracted chiral condensate normalized to the corresponding 
zero temperature value (right). In the right hand figure data for $N_\tau =6$ 
calculations have been 
shifted by $-7$~MeV (asqtad) and $-5$~MeV (p4),
}
\label{fig:condensate}
\end{figure}

At zero quark mass, the chiral condensate needs to be 
renormalized only multiplicatively.
At nonzero values of the quark mass, an additional renormalization 
is necessary to eliminate singularities that are proportional to $m_q/a^2$.
An appropriate observable that takes care of the additive renormalizations is 
obtained by subtracting 
a fraction, proportional to $m_l/m_s$, of the strange quark condensate from the light quark condensate.
To remove the multiplicative renormalization factor we divide this difference at finite temperature by 
the corresponding zero temperature difference, calculated at the same value
of the lattice cutoff, $i.e.$,
\begin{equation}
\Delta_{l,s}(T) = \frac{\langle \bar{\psi}\psi \rangle_{l,T} -
\frac{m_l}{m_s}
\langle \bar{\psi}\psi \rangle_{s,T}}{\langle \bar{\psi}\psi \rangle_{l,0} -
\frac{m_l}{m_s} \langle \bar{\psi}\psi \rangle_{s,0}} \; .
\label{delta_ls}
\end{equation}
This observable has a sensible chiral limit and is 
an order parameter for chiral symmetry breaking. It is unity at low 
temperatures and vanishes at $T_c$ for $m_l=0$. For the quark mass 
values used in our study of bulk thermodynamics, 
{\it i.e.}, $m_l=0.1 m_s$, its temperature dependence is shown 
in Fig.~\ref{fig:condensate}. It is evident that $\Delta_{l,s}(T)$ varies
rapidly in the same temperature range as the bulk thermodynamic
observables and the light quark number susceptibility. Based on 
this agreement we conclude that the onset of liberation of light quark and
gluon degrees of freedom (deconfinement) and chiral symmetry restoration 
occur in the
same temperature range in QCD with almost physical values of the
quark masses, {\it i.e.}, in a region of the QCD phase diagram where the
transition is not a true phase transition but rather a rapid crossover.

Cutoff effects in the chiral condensate as well as in bulk thermodynamic
observables are visible when comparing calculations performed with
a given action at two different
values of the lattice cutoff, e.g., $N_\tau=6$ and $8$. These cutoff
effects can to a large extent be absorbed in a common shift of the 
temperature scale. This is easily seen by comparing the left and right
panels in  Fig.~\ref{fig:condensate}. A global shift of the temperature 
scale used for the $N_\tau=6$ data sets by $5$~MeV for the p4 and by $7$~MeV 
for the asqtad action makes the $N_\tau=6$ and $8$ data sets coincide 
almost perfectly. 
This is similar in magnitude to the cutoff dependence observed in 
$(\epsilon-3p)/T^4$ and again seems to reflect the 
cutoff dependence of the transition temperature as well as 
residual cutoff dependencies of the zero
temperature observables used to determine the temperature scale. 

As can be seen from Fig.~\ref{fig:condensate}(right), even after
the shift of temperature scales the chiral condensates calculated with 
asqtad and p4 actions show
significant differences. This reflects cutoff effects arising
from the use of different discretization schemes and to some extent is 
also due to the somewhat different physical quark mass values on the
LCPs for the asqtad and p4 actions. The differences are most
significant at low temperatures where the lattice spacing is quite 
large. Here, cutoff effects that arise from the explicit breaking of flavor 
symmetry in the staggered fermion formulations may become important.
At temperatures larger than the crossover temperature, the chiral condensate
obtained from calculations with the asqtad action is systematically larger 
than that obtained with the p4 action. This is consistent with the fact
that the quark masses used on the asqtad LCP are somewhat larger than
those on the p4 LCP. 

We note that at finite temperatures the chiral condensate
as well as $\Delta_{l,s}(T)$ are quite sensitive to the quark mass. In the
chiral symmetry broken phase, in both cases the leading order quark mass 
correction  is proportional to the square root of the light
quark mass \cite{goldstone}. Small differences in the actual quark mass values used 
in p4 and asqtad calculations on lattices with temporal extent $N_\tau =6$ and $8$ will thus be
enhanced in the transition region. This makes it important to have good control of the
line of constant physics.
We will discuss the quark mass dependence of the chiral condensate and quark number 
susceptibilities as well as the cutoff dependence of pseudo-critical temperatures
extracted from them in more detail in a forthcoming publication \cite{hotTc}. 

\subsection{The Polyakov loop}

The logarithm of the Polyakov loop is 
related to the change in free energy induced by a static quark source.
It is a genuine order parameter for deconfinement only for the 
pure gauge theory, $i.e.$, all quark masses taken to infinity. 
At finite quark masses 
it is nonzero at all values of the temperature but changes rapidly at the 
transition. 
The Polyakov loop operator is not present in the QCD action but can be 
added to it as an external
source. Its expectation value is then given by the derivative of the
logarithm of the modified partition function
with respect to the corresponding coupling, evaluated at zero 
coupling. As far as we know, the Polyakov loop
is not directly sensitive to the singular structure
of the partition function in the chiral limit. Therefore, its 
susceptibility will not diverge at $\bar{m}=0$ nor is its slope in the 
transition region related to
any of the critical exponents of the chiral transition. 
Nonetheless, the Polyakov loop is observed to vary rapidly in the transition 
region indicating that the screening of static quarks  
suddenly becomes more effective. This in turn leads to a reduction of the
free energy of static quarks in the high temperature phase of QCD. 

\begin{figure}[t]
\begin{center}
\epsfig{file=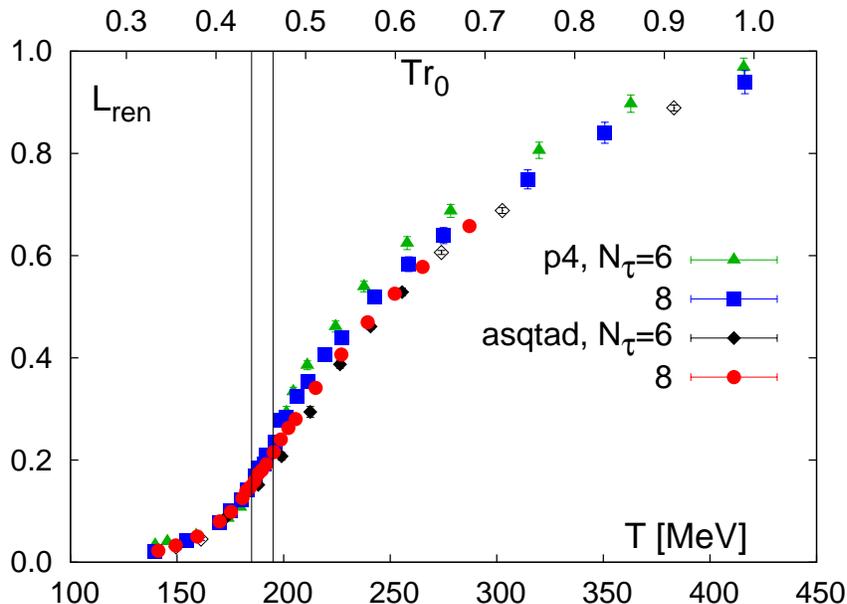,width=12.0cm}
\end{center}
\caption{\label{fig:Polyakov}(color online) The renormalized Polyakov loop obtained
with the asqtad and p4 actions from simulations on lattices with 
temporal extent $N_\tau =6$ and $8$.
Open symbols for the
$N_\tau=6$, asqtad data set denote data obtained with the R~algorithm. All
other data have been obtained with an RHMC algorithm.
}
\end{figure}

The Polyakov loop needs to be renormalized in order to eliminate
self-energy contributions to the static quark free energy. 
For the p4 action, this renormalization factor is obtained 
from the renormalization of the heavy quark potential as outlined in 
Ref.~\cite{rbcBIeos}. 
In calculations with the asqtad action, we apply the same 
renormalization procedure and details of this 
calculation are given in Appendix B. 
The results for the renormalized operator for both actions 
are shown in Fig.~\ref{fig:Polyakov}. Similar to other observables 
discussed in this paper, we also observe for the Polyakov loop expectation
value that results obtained on lattices with temporal extent $N_\tau=6$
are shifted relative to data obtained on the $N_\tau=8$ lattices by
about $5$~MeV. The renormalized Polyakov loop rises significantly in the
transition region. The change in slope, however, occurs in a rather broad
temperature interval. Similar to the strange quark number susceptibility
the Polyakov loop does not seem to be well suited for a quantitative
characterization of the QCD transition, as it is not directly related
to any derivatives of the singular part of the QCD partition function.
In fact, even in the chiral limit 
we do not expect that $L_{\rm ren}$ or its susceptibility will
show pronounced critical behavior.

\section{Conclusions}

We have presented new results on the equation of state of QCD with
a strange quark mass chosen close to its physical value
and two degenerate light
quarks with one tenth of the strange quark mass. A comparison of 
calculations performed with the p4 and asqtad staggered fermion 
discretization schemes shows that both actions lead to a consistent
picture for the temperature dependence of bulk thermodynamic observables
as well as other observables that characterize deconfining and
chiral symmetry restoring aspects of QCD thermodynamics. The calculations
performed on lattices with temporal extent $N_\tau=8$ suggest that 
the deconfinement of light partonic degrees of freedom, which is reflected
in the rapid change of the energy density as well as the light quark
number susceptibility, goes along with a sudden drop in the chiral 
condensates, indicating the melting of the chiral condensate. These
findings confirm earlier results obtained within these two discretization 
schemes.

A comparison of results obtained with the asqtad and p4 actions for two 
different values of the cutoff, $aT=1/6$ and $1/8$, suggests  that
cutoff effects in both discretization schemes are at most a few percent for 
temperatures larger than $300$~MeV. In the vicinity of the peak in the 
trace anomaly, cutoff effects are still about 15\% in calculations with the
p4 action and at most half that size for the asqtad action. 
Also, at low temperatures both actions give consistent results. Here, however,
further studies with lighter quarks on finer lattices are needed to 
get better control over cutoff effects that distorted the hadron spectrum 
and might influence the thermodynamics in the chiral symmetry broken
phase.

We find that different observables give a consistent characterization of
the transition region, $180$~MeV~$\lsim T \lsim$~$200$~MeV. 
A comparison of results obtained on lattices with temporal extent 
$N_\tau=6$ and $8$ shows that with decreasing lattice spacing the 
transition region shifts by about 5--7~MeV towards smaller values of 
the temperature. Assuming that cutoff effect are indeed ${\cal O}(a^2)$ 
for our current simulation parameters, one may expect a shift of similar
magnitude when extrapolating to the continuum limit. Preliminary studies
of the quark mass dependence of the transition region  \cite{Soeldner} 
suggest that a 
shift of similar magnitude is to be expected when the ratio of light
to strange quark masses is further reduced to its physical value,
$m_s/m_q \simeq 25$. A more detailed
analysis of the transition region, its quark mass and cutoff dependence, 
will be discussed in a forthcoming publication \cite{hotTc}.

\section*{Acknowledgments}
\label{ackn}
This work has been supported in part by contracts DE-AC02-98CH10886, DE-AC52-07NA27344,
DE-FG02-92ER40699, DE-FG02-91ER-40628, DE-FG02-91ER-40661, DE-FG02-04ER-41298,
DE-KA-14-01-02 with 
the U.S. Department of Energy, and NSF grants PHY08-57333, 
PHY07-57035, PHY07-57333 and PHY07-03296,
the Bundesministerium f\"ur Bildung und Forschung under grant
06BI401, the Gesellschaft
f\"ur Schwerionenforschung under grant BILAER and the Deutsche
Forschungsgemeinschaft under grant GRK 881.
We thank the computer support staff for the BlueGene/L computers at at
Lawrence
Livermore National Laboratory (LLNL) and the New York Center for
Computational
Sciences (NYCCS), on which the numerical simulations have been performed.
\appendix

\section{EoS with improved staggered fermion actions}

In this appendix we review and collect details of the p4 \cite{Heller} and asqtad \cite{Orginos} 
staggered improved actions to present a unified framework for the description of thermodynamic calculations 
presented in this paper. 

\subsection{The asqtad and p4 actions}

The QCD partition function on a lattice of size
$N_\sigma^3 N_\tau$ is written as 
\beqn
Z_{LCP}(\beta, N_\sigma, N_\tau) =  \int \prod_{x,\mu} {\rm d}U_{x,\mu}
 {\rm e}^{- S(U)} \;\; ,
\label{partition}
\eqn
where $U_{x,\mu}\in SU(3)$ denotes the gauge link variables, and
\begin{equation}
S(U) = \beta S_G(U,u_0(\beta)) - S_F(U,u_0(\beta)) 
\label{action}
\end{equation}
is the Euclidean action. We define the tadpole coefficient $u_0$ as 
$u_0=\langle P\rangle_0^{1/4}$ where $P$ is 
the $1 \times 1$ Wilson loop called 
the plaquette and defined below in Eq.~(\ref{PRC}). 
After integration over the fermion fields the QCD action is written as the sum of a purely gluonic contribution, $S_G$, and the
fermionic contribution, $S_F$, involving only the gauge fields
\begin{eqnarray}
S_F(U,u_0(\beta)) &=& \frac{1}{2} {\rm Tr}\ln  D(u_0(\beta),\hat{m}_l(\beta))
+ \frac{1}{4} {\rm Tr}\ln  D(u_0(\beta),\hat{m}_s(\beta)) \;\; .
\label{SF}
\end{eqnarray}
Here the factors $1/2$ and $1/4$ arise from taking the square root and the fourth root of the staggered fermion determinant to
represent the contribution of 2 degenerate light flavors and the single strange flavor to the QCD
partition function \cite{root}.
We show the explicit lattice-coordinate components of the Dirac
operator (fermion matrix) and write it in terms of its diagonal and
off-diagonal contributions,
\begin{equation}
D_{xy}(u_0(\beta),\hat{m}_i(\beta)) = \hm_i(\beta) \delta_{xy} +M_{xy}(u_0(\beta)) \; .
\label{fmatrix}
\end{equation}
We also introduce the shorthand notation
$D_i = \hm_i + M = D(u_0(\beta),\hat{m}_i(\beta))$.

In Eqs.~(\ref{action}) to (\ref{fmatrix}) we made explicit the dependence on 
$\beta$, $i.e.$ in addition to 
the explicit multiplicative $\beta$-dependence in front of $S_G$ the  
actions also depend on $\beta$ through the tadpole factor $u_0$.

The general form of the gluonic part of the action as it is used in the asqtad 
and p4 actions is  given in terms of a 4-link
plaquette term (pl), a planar 6-link rectangle (rt) and a twisted 6-link loop (pg),
\begin{equation}
S_G = \sum_{x,\mu < \nu} \left( \beta_{\rm pl} (1-P_{\mu\nu}) +
\beta_{\rm rt} (1-R_{\mu\nu})\right) +
\beta_{\rm pg} \sum_{x,\mu < \nu<\sigma} (1-C_{\mu\nu\sigma}) 
\; ,
\label{action_asqtad_G}
\end{equation}
with
$\mu,\; \nu,\; \sigma \in 1,\;2,\;3,\;4$ and $P_{\mu\nu}$, $R_{\mu\nu}$ and 
$C_{\mu\nu\sigma}$ denote 
the normalized trace of products of gauge field variables $U_{x,\mu}$:
\begin{eqnarray}
P_{\mu\nu} &=& \frac{1}{3}{\rm Re\; Tr}\; U_{x,\mu} U_{x+\hat{\mu},\nu} U^\dagger_{x+\hat{\nu},\mu}
U^\dagger_{x,\nu} \nonumber \\
R_{\mu\nu} &=& \frac{1}{6}{\rm Re\; Tr}\left( U_{x,\mu} U_{x+\hat{\mu},\mu} U_{x+2\hat{\mu},\nu}
U^\dagger_{x+\hat{\mu}+\hat{\nu},\mu} U^\dagger_{x+\hat{\nu},\mu} U^\dagger_{x,\nu} + (\mu \leftrightarrow \nu)
\right) \nonumber \\
C_{\mu\nu\sigma} &=& \frac{1}{12}
{\rm Re\; Tr}\left( 
U_{x,\mu} U_{x+\hat{\mu},\nu} U_{x+\hat{\mu}+\hat{\nu},\sigma}
U^\dagger_{x+\hat{\nu}+\hat{\sigma},\mu} U^\dagger_{x+\hat{\sigma},\nu}
U^\dagger_{x,\sigma} \right. \nonumber \\
&&\hspace*{1.55cm}\left. + (\mu \leftrightarrow \nu) + (\nu \leftrightarrow \sigma)
+(\mu \leftrightarrow \nu,\;\mu \leftrightarrow -\mu)
\right) \; ,
\label{PRC}
\end{eqnarray}
where $\hat{\mu}$ is the unit vector in $\mu$-direction.
The p4 action only contains the first two terms, while the asqtad action
contains all three terms with $\beta$-dependent couplings. The set of couplings
for both actions is given in Table~\ref{tab:pg_couplings}.
\begin{table}[htb]
\begin{center}
\begin{tabular}{|c|c|c|c|c|c|c|c|c|}
\hline
~&$\beta_{pl}$ & $\beta_{\rm rt}$ & $\beta_{\rm pg}$ & $\beta'_{\rm rt}$ &
$\beta'_{\rm pg}$ & $c_1$ & $c_3$ & $c_{12}$ \\[2mm]
\hline
& & & & & &  &  & \\[-4mm]
p4 & $\displaystyle{\frac{5}{3}}$ & $\displaystyle{-\frac{1}{6}}$ & 0 & 0& 0 & $\displaystyle{\frac{3}{8}}$
& 0 & $\displaystyle{\frac{1}{96}}$\\[2mm]
\hline
& & & & & &  &  & \\[-4mm]
asqtad & 1 &
$\displaystyle{-\frac{1}{10u_0^2}(1+ 0.4805 \alpha_s)}$ &
$\displaystyle{-0.1330 \frac{1}{u_0^2} \alpha_s}$ &
$\displaystyle{\frac{1}{5u_0^3} \left(1.3132+ 0.4805 \alpha_s \right)}$  &
$\displaystyle{\frac{1}{u_0^3} \left( 0.1734+ 0.2660 \alpha_s \right)}$  &  $\displaystyle \frac{1}{2}$ &  $\displaystyle -\frac{1}{48 u_0^2}$ & $\displaystyle 0$
\\[3mm]
\hline
\end{tabular}
\end{center}
\caption{
Couplings $\beta_x$, $x=$ pl, rt, pg defining the pure gauge part of the p4 and 
asqtad actions, respectively. Here $u_0$  is the tadpole coefficient and 
$\alpha_s = -4 \ln (u_0) / 3.0684$. Also given are the derivatives 
$\beta'_x(u_0) = {\rm d}\beta_x/{\rm d} u_0$ and the values of the coefficients 
$c_i$ which appear in the fermion part of the action.
}
\label{tab:pg_couplings}
\end{table}
We note here a difference in convention that exists between the asqtad and p4
actions. In the asqtad action, the full constant factor in front of the plaquette term
is defined to be $\beta$. In the continuum limit,
its relation to the gauge 
coupling $g^2$ is $\beta = 10/g^2$. In the p4 action, the convention
of the standard Wilson plaquette action has been kept, $\beta = 6/g^2$. For this
reason the plaquette term contains an extra factor $\beta_{\rm pl}=5/3$.
Also note that
the pure gauge part of the asqtad action reduces to that of the p4 action
for $u_0=1$. 

The fermionic part of the action has been improved by introducing higher order
difference schemes (one-link and three-link terms) to discretize the derivative in the 
kinetic part of the action \cite{Nai89},
\begin{eqnarray}
D[U]_{ij}&=& m~\delta_{ij} + \Big(c_{1} ~{ A[U]_{ij}} 
+ c_{3} ~{ B_1[U]_{ij}}+  c_{12} ~{ B_2[U]_{ij}} \Big)
\label{fmatrix2}
\\[2mm]
{A[U]_{ij}}
&=&\apart{i}{j}{\mu} \nonumber \\
{B_1[U]_{ij}}
&=&\bparta{i}{j}{\mu} \nonumber 
\end{eqnarray}
\vspace*{-0.5cm}
\begin{eqnarray*}
B_2[U]_{ij} &=&
\bpartb{i}{j}{\mu}{\nu}\; .\\
\end{eqnarray*}
Here $\eta_\mu(i)$ denotes the staggered phase factor. 
The coefficients $c_1$, $c_3$ and $c_{12}$
for p4 and asqtad actions are also given in Table \ref{tab:pg_couplings}.
The overall normalization is such that in the limit
$\beta\rightarrow \infty$ we recover the naive continuum
action for Dirac fermions.
In both actions, so-called fat links ($U^{\rm fat}$) have been introduced in the 
1-link terms; {\it i.e.}, in addition to the straight 1-link parallel transporter 
$U_{x,\mu}$ that connects adjacent sites $(x,x+\hat{\mu})$, additional
longer paths have been added. In the case of the p4 action,
this is a simple three-link path, called the staple, going around an elementary plaquette. 
The coefficients of the 1-link term and the staple are $\beta$-independent
and equal to $1/(1+6 \omega)$ and $\omega/(1+6 \omega)$, respectively. We used  $\omega=0.2$ 
\cite{Peikert}.
The asqtad fat-link contains additional non-planar 5-link and 7-link paths to remove completely
the effect of flavor symmetry breaking to order $(g^2 a^2)$ \cite{Orginos}
as well as a planar 5-link path,
the so-called Lepage term. The coefficient of the one link term in this case is equal to $5/8$.
The coefficients of the 3-, 5- and 7-link paths are equal
to $1/(2 u_0^{2})(1/8),~1/(8 u_0^{4})(1/8)$ and $1/(48 u_0^{6})(1/8)$, respectively. 
Furthermore, the coefficient of the Lepage term is $-1/(16 u_0^4)$.

Finally, we point out that the conventions for the asqtad fermion 
matrix given in \cite{Orginos1} differ from those introduced in 
Eqs.~(\ref{fmatrix}) and (\ref{fmatrix2}) by an 
overall factor two. Thus, all couplings given in \cite{Orginos} need to be divided 
by a factor two to match the expressions given here.

\subsection{\boldmath $\beta$-functions}

In this work,
we consider the thermodynamics of ($2+1$)-flavor QCD along a {\it line of constant
physics} (LCP). The LCP is fixed by choosing two degenerate light ($\hm_l$) and a 
heavier (strange)
quark mass ($\hm_s$) as functions of the gauge coupling $\beta$, i.e. $\hm_{l,s}\equiv
\hm_{l,s}(\beta)$, such that physical observables, e.g. a set of hadron masses,
calculated at $T=0$ at the same value of the cutoff as used at finite temperature,
stay constant. In calculations with the asqtad as well with the p4 action, it 
turns out that a LCP is well characterized by specifying $\hm_l(\beta)$ and keeping
the ratio $\hm_l/\hm_s$ fixed. Explicit parametrizations for $\hm_l(\beta)$ have been 
given for
the asqtad \cite{milc_eos} and p4 \cite{rbcBIeos} actions for $\hm_l/\hm_s =0.1$ 
in the parameter range relevant for the thermodynamic calculations discussed here.

The basic observable we need to calculate is the trace anomaly, $\epsilon -3p$, 
introduced in Eq.~(\ref{delta}). To do so, we need to relate temperature changes 
to changes of the gauge coupling. These are controlled by the $\beta$-function,
\begin{equation}
R_\beta = T \frac{{\rm d} \beta}{{\rm d}T} = - a \frac{{\rm d} \beta}{{\rm d}a} \; .
\label{Rb}
\end{equation}
The $\beta$-function can be determined by analyzing the $\beta$-dependence of a physical
observable expressed in lattice units. We use here distance scales $r_n$ extracted
from the heavy quark potential as introduced in Eq.~(\ref{r0r1}).
Explicit parametrizations for $\hat{r}_0 (\beta)$ (p4-action) and $\hat{r}_1(\beta)$ 
(asqtad
action) have been given in \cite{rbcBIeos} and \cite{milc_eos}, respectively.
Using the explicit parametrizations for $\hat{r}_n$ and $\hm_l$ we can determine the
$\beta$-function $R_\beta$ and the mass renormalization function $R_m$,
\beqn
R_\beta (\beta) =  \hat{r}_n \left(
\frac{{\rm d} \hat{r}_n}{{\rm d}\beta} \right)^{-1}
\; , \; 
R_m(\beta) = \frac{1}{\hm_l(\beta)}
\frac{{\rm d} \hm_l(\beta)}{{\rm d}\beta}
\; .
\label{rnbeta}
\eqn
In the case of the asqtad action one more function plays a central role
in the derivation of general expressions for thermodynamic quantities. This
is the tadpole coefficient $u_0 (\beta)$ which enters the definition of the asqtad
action. It is given in terms of the plaquette expectation value at zero temperature, 
$u_0=\langle P \rangle_0^{1/4}$, where $P$ 
denotes the product of gauge field variables defined on an elementary plaquette of the 
4-dimensional lattice. It is introduced in Eq.~(\ref{PRC}).
As discussed previously, the p4 action contains only tree-level improved terms without tadpole 
improvement. In order to keep the following formulas 
valid for both actions, we insert $u_0 \equiv 1$ as a trivial constant in the p4 action.

Since $u_0$ depends on $\beta$ for the asqtad action, we also
need it's derivative with respect to $\beta$,
\begin{equation}
R_u(\beta) = \beta \frac{{\rm d} u_0(\beta)}{{\rm d}\beta} \; .
\label{Ru}
\end{equation}
This is obtained from the polynomial parametrization used for interpolating 
the value of the tadpole factor in the asqtad action,
\begin{equation}
u_0(\beta) = a_0 + a_1 b + a_2 b^2  + a_3 b^3 + a_4 b^4 + a_5 b^5 + a_6 b^6
\; ,
\label{tadpole}
\end{equation}
with $b=\beta - 6.60$, $a_0 = 0.86158$, $a_1 = 0.0426043$, $a_2 =
-0.0254633$, $a_3 = 0.0261288$, $a_4 = -0.0116944$, $a_5 = -0.0417343$,
and $a_6 = 0.0436528$.
This parametrization is suitable for the  $\beta$-range covered by the calculations presented
in this work. In the weak coupling limit $R_u \sim \beta^{-1}$.

\subsection{Thermodynamics}
Having established the notation, we now present expressions for basic thermodynamic quantities 
calculated with the asqtad and p4 actions. 

The light and strange quark condensates calculated
at finite ($x=\tau$) and zero ($x=0$) temperature are 
\beqn
\langle\bar{\psi}\psi\rangle_{q,x} \equiv \frac{1}{4}
\frac{1}{N_\sigma^3N_x} \left\langle {\rm Tr} D_q^{-1}
\right\rangle_x \;\; ,\;\;
q=l,~s\;\; , \;\; x=0,~\tau \;\; .
\label{quarkcondensate}
\eqn
Here $N_x$, with $x=0,\; \tau$ denotes the temporal extent of zero and finite 
temperature lattices, and $N_\sigma$ is the size in the spatial directions. 
The action densities are then given by 
\beqn
s \equiv \frac{1}{N_\sigma^3 N_x} S  \;\;{\rm for}\;\; x=0,~\tau
\label{gluondensity}
\eqn
and similarly for $S_G$ and $S_F$. We introduce a short-hand notation
for differences of expectation values of intensive observables calculated at 
finite and zero temperature as 
\begin{equation}
\Delta\VEV{X} = \VEV{X}_0 -\VEV{X}_\tau \; .
\label{dif}
\end{equation}

The trace anomaly is then given by 
\begin{eqnarray}
\frac{\epsilon-3p}{T^4} =
\frac{\Theta^{\mu\mu} (T)}{T^4}
&=&  R_\beta(\beta) N_\tau^4 \Delta\VEV{s} \; ,
\label{deltaLGT}
\end{eqnarray}
and Eq.~(\ref{deltaLGT}) can be rewritten as 
\begin{eqnarray}
\frac{\Theta^{\mu\mu} (T)}{T^4} &=&
\frac{\Theta^{\mu\mu}_G(T)}{T^4} +
\frac{\Theta^{\mu\mu}_F(T)}{T^4} \;\; ,
\label{e3p}
\end{eqnarray}
with $\Theta^{\mu\mu}_F(T)/T^4$ denoting the contribution from the
renormalization group invariant contribution of 
light and strange quark condensates,
\begin{eqnarray}
\frac{\Theta^{\mu\mu}_F(T)}{T^4}  
&=&-R_\beta R_m N_\tau^4  
\left( 2 \hm_l \Delta\VEV{\bar \psi \psi}_l
        + \hm_s \Delta\VEV{\bar \psi \psi}_s \right) \; ,
\label{e3pfermion}
\end{eqnarray}
and ${\Theta^{\mu\mu}_G(T)/T^4}$ including all the remaining terms
that would survive the chiral limit ($m_{l,s}\rightarrow 0$),
\begin{eqnarray}
\frac{\Theta^{\mu\mu}_G(T)}{T^4}  
&=&  R_\beta N_\tau^4 \left( \Delta\VEV{s_G} -R_u 
\left( 
6 \beta'_{\rm rt} \Delta\VEV{R} + 4 \beta'_{\rm pg} \Delta\VEV{C}
+\frac{1}{4\beta}  
\Delta\VEV{{\rm Tr}\left(\left(2 D_l^{-1}+D_s^{-1}\right)
\frac{d M}{d u_0} \right) }\right)
\right)
 \; . 
\label{e3pgluon} 
\end{eqnarray}
Here $\VEV{s_G}$ denotes the contribution of the gluonic action density and
$\VEV{R}$ and $\VEV{C}$ denote the expectation values of 
6-link Wilson loops introduced in Eq.~(\ref{PRC}).
The functions $\beta'_{\rm rt}$ and $\beta'_{\rm pg}$ are given in 
Table~\ref{tab:pg_couplings}. We have explicitly separated contributions
proportional to the derivative of the tadpole coefficient, $R_u$. 
As $R_u\equiv 0$ for the tree level improved p4 action, these terms contribute 
only in calculations with the
asqtad action, where they help to reduce the cutoff dependence of thermodynamic
observables at nonzero lattice spacing. 

In Appendix D we also use the obvious short hand notation for light and 
strange fermion contributions, 
$\displaystyle{\Theta^{\mu\mu}_F(T)/T^4 =\Theta^{\mu\mu}_{F,l}(T)/T^4
+\Theta^{\mu\mu}_{F,s}(T)/T^4}$.

\section{Polyakov loop renormalization for the asqtad action}

\begin{figure}
\centering
\includegraphics[width=0.5\textwidth]{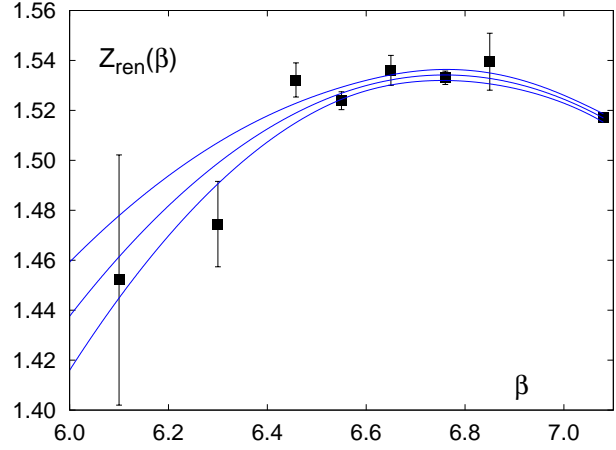}
\caption{(color online) The renormalization constant $Z(\beta)$. The error band
is determined with bootstrap analysis.}
\label{Zbeta}
\end{figure}

To renormalize the Polyakov loop we use an approach similar to the 
one of Ref.~\cite{rbcBIeos}: we renormalize the static quark
potential and extract the self-energy of a static quark. At a given
lattice spacing the latter is simply a constant shift of the potential
which we denote $C(\beta)$. As a matching condition we have
chosen the same one as for the p4 action: the renormalized potential
is set equal to the string potential 
$V_{string}(r)=-\pi/12r+\sigma r$ at distance $r=1.5r_0$.
We first fit the static potential measured from the ratios of
the correlators of Wilson lines in the Coulomb gauge to
the ansatz:
\begin{equation}
  V(r,\beta)=V_0-\frac{1.65-\sigma r_0^2}{r}+\sigma r
\end{equation}
with $V_0$, $\sigma$, $r_0$ being the fit parameters,
and then apply the conditions:
\begin{eqnarray}
  V(r,\beta) &=& V_{ren}(r,\beta)+C(\beta),\\
  V_{ren}(r=1.5r_0,\beta) &=& V_{string}(r=1.5r_0).
\end{eqnarray}
This gives
\begin{equation}
  C(\beta)=V_0+\frac{1}{1.5r_0}\left(
  \frac{\pi}{12}-1.65+\sigma r_0^2\right).
\end{equation}
The results for $C(\beta)$ are collected in 
Table~\ref{tabC}. The renormalization of the Polyakov loop
amounts to removing the self-energy contribution
\begin{equation}
  L_{ren}(\beta)=L(\beta)e^{(C(\beta)/a)/(2T)}=L(\beta)
  \left(Z(\beta)\right)^{N_\tau},
\end{equation}
where we defined
\begin{equation}
  Z(\beta)=e^{C(\beta)/2}
\end{equation}
for convenience. In Fig.~\ref{Zbeta}, $Z(\beta)$ data is shown
together with the fit to a quadratic polynomial. The error band
is estimated with a bootstrap analysis.

\begin{table}
\centering
\begin{tabular}{|l|l|}
\hline
$\beta$ & $C(\beta)$ \\ \hline
6.100 & 0.746(69) \\
6.300 & 0.777(23) \\
6.458 & 0.8534(89) \\
6.550 & 0.8425(47) \\
6.650 & 0.8584(78) \\
6.760 & 0.8545(35) \\
6.850 & 0.863(15) \\
7.080 & 0.8335(22) \\
\hline
\end{tabular}
\caption{Twice the self energy of a static quark $C(\beta)$ in
lattice units along the line of constant physics $m_l=0.1m_s$.}
\label{tabC}
\end{table}

\section{Parametrizing the Equation of State for  Hydrodynamics}
\label{sec:hydro}

Here we present a simple functional form for generating the equation of state that can be readily applied to most hydrodynamic models of heavy ion collisions.  The equation of state is an essential input for solving the hydrodynamic equations of motion, but as explained in Section~\ref{sec:trace_anomaly}, some form of interpolation of the interaction measure is required to generate the equation of state from the lattice data.  However, small fluctuations in the equation of state can be magnified through the time evolution in the hydrodynamic models and can lead to anomalous effects in the final state.  Furthermore, it is preferable for the equation of state to transition smoothly to the EoS of the hadronic resonance gas that is imposed when the models freeze-out to produce final state hadrons.  Most hydrodynamic models use a simplified version of the equation of state that incorporates a hadronic resonance gas EoS below the transition.  Near and above the transition some models approximates the crossover EoS calculated on the lattice~\cite{Huovinen:2005gy,Luzum:2008cw,Pratt:2008qv}, but it is more common for hydro models to rely on a bag model equation of state with a first order phase transition~\cite{Kolb:2002ve,Hirano:2005wx,Nonaka:2007bj}.

\begin{figure}[thb]
\begin{center}

\epsfig{file=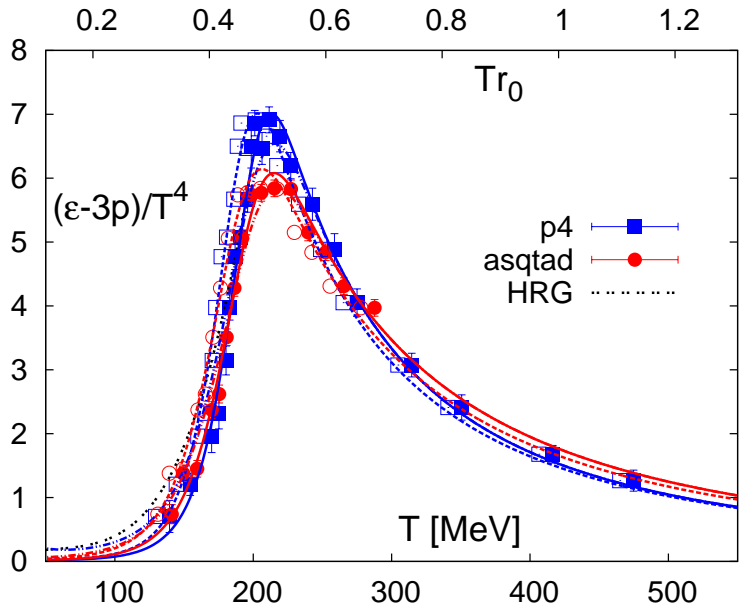,width=12.0cm}
\caption{(color online) Fits to the trace anomaly using Eq.~(\ref{eq:simplefit}) for p4 and asqtad $N_\tau=8$ lattice data (solid lines), to lattice data merged with the hadron resonance gas calculation ($m_{res}<2.5$~GeV) over the region $100 < T < 130$~MeV (double-dot dashed), and to the lattice data shifted to lower temperature by 10~MeV (dashed).}
\label{fig:simplefit}
\end{center}
\end{figure}
To remedy this gap between theory and phenomenology we fit the lattice calculation of the trace anomaly for both p4 and asqtad actions to the simple functional form given by Eq.~(\ref{eq:simplefit}).  
\begin{equation}
\frac{\epsilon -3p}{T^4} = 
\left( 1 - \frac{1}{[1+e^{(T-c_1)/c_2)}]^{2}} \right)
\left(  \frac{d_2}{T^2}+  \frac{d_4}{T^4} \right)
\label{eq:simplefit} 
\end{equation}
This form uses a modified hyperbolic tangent to describe the transition region and retains the high temperature region {\em ansatz} of Eq.~(\ref{e3phigh}).  For the p4 action, the high temperature parameters are set to the values obtained in Section~\ref{sec:highT} for the range $T>0.25$~GeV.  We repeat this procedure for the asqtad action, but with the fourth order term $d_4$ set to zero due to the lack of high temperature measurements to constrain it.  The high temperature parameters are then fixed as the full function of Eq.~(\ref{eq:simplefit}) is fit to the lattice data or to some combination of lattice data and hadronic resonance gas.  In this way we prevent any deficiencies in the description of the transition and peak region from biasing the high temperature behavior, although variations at low temperature will produce offsets in the pressure and energy density at high temperature.  To match to the HRG EoS we adopt a procedure that is similar to systematic error analysis of Section~\ref{sec:thermodynamics} in which the HRG was used establish the starting value for the integration of the pressure at 100~MeV.  However, to achieve a gradual transition we incorporate the HRG calculation for $m_{res}<2.5$~GeV over a range of temperatures $100 < T < 130$~MeV.  HRG values for the trace anomaly are sampled every 5~MeV and assigned an error of 0.1 to produce weights that are comparable to the low temperature lattice data.

Fig.~\ref{fig:simplefit} shows the result of fitting Eq.~(\ref{eq:simplefit}) to the p4 and asqtad $N_\tau$ trace anomaly (solid lines).  The trace anomaly for the resonance gas is also plotted (double dotted), along with fits to the both p4 and asqtad data that are combined with the HRG for $100 < T < 130$~MeV.
As an additional test of this parametrization, we also fit to the lattice data shifted to lower temperature by 10~MeV (open symbols).  Based on the comparison to $N_\tau=6$ shown in Fig.~\ref{fig:details}, we expect that the continuum extrapolation will lead to shifts that are no greater than this amount.  These fits follow the same prescription: the high temperature component is fit first and these parameters are fixed for the full minimization (dashed lines).

\begin{table}[h]
\begin{center}
\vspace{0.3cm}
\begin{tabular}{|l|c|c|c|c|c|}
\hline
Data & $d_2$ [GeV]$^2$ & $d_4$ [GeV]$^4$ & $c_1$ [GeV] & $c_2$ [GeV] &
$\chi^2$/dof \\
\hline
            p4  & 0.24(2) &  0.0054(17) &  0.2038(6)   &  0.0136(4)   
&  26.7/19 \\
p4-10~MeV   & 0.241(6) &  0.0035(9) &  0.1938(6) &  0.01361(4) & 26.7/19 \\
   HRG+p4  & 0.24(2) &  0.0054(17) &  0.2073(6)   &  0.0172(3)   &  
-- \\
\hline
              asq   & 0.312(5)  & 0.00  &  0.2024(6)  &  0.0162(4) & 
34.4/14 \\
asq-10~MeV & 0.293(6) &  0.00 &  0.1943(6) &  0.01670(4) & 42.8/14 \\
HRG+asqtad & 0.312(5) & 0.00 &  0.2048(6)  &  0.0188(4) & -- \\
\hline
\end{tabular}
\caption{Parameter values for fits of Eq.~(\ref{eq:simplefit}) to trace
anomaly data for p4 and asqtad, data combined with HRG calculations, and
data shifted by 10~MeV.}
\label{tab:simplefit}
\end{center}
\end{table}
The parameters for all fits are  listed in Tab.~\ref{tab:simplefit}.  As is readily seen in Fig.~\ref{fig:simplefit} and the $\chi^2/ndf$ reported for the lattice fits in Tab.~\ref{tab:simplefit}, this functional form provides only an approximation to the full lattice calculation, but one that will be shown to be within the systematic errors for the equation of state.  As statistical and systematic errors are reduced in future calculations, the parametrization of Eq.~(\ref{eq:simplefit}) is easily modified to include additional terms, including the high temperature perturbative terms.  The shift by 10~MeV has the predictable effect of lowering the $c_1$ parameter by a similar amount.  Including the HRG points affects mainly the exponential slope term $c_2$ leading to a slight reduction of the peak.

The energy density and pressure are calculated by numerically integrating the trace anomaly fits according to Eq.~(\ref{pres}).  For these parametrizations the integration is started at 50~MeV. Because these parametrizations of the trace anomaly have their minima in this region, the pressure and energy density are not sensitive to the exact location of the starting temperature.  We note, however, that Eq.~(\ref{fig:simplefit}) rises rapidly as the temperature is further reduced, and is not suitable for extrapolating to temperatures less than 50~MeV, well below the freeze-out temperature for all hydrodynamic calculations for relativistic heavy ion collisions.

\begin{figure}[t]
\begin{center}
\epsfig{file=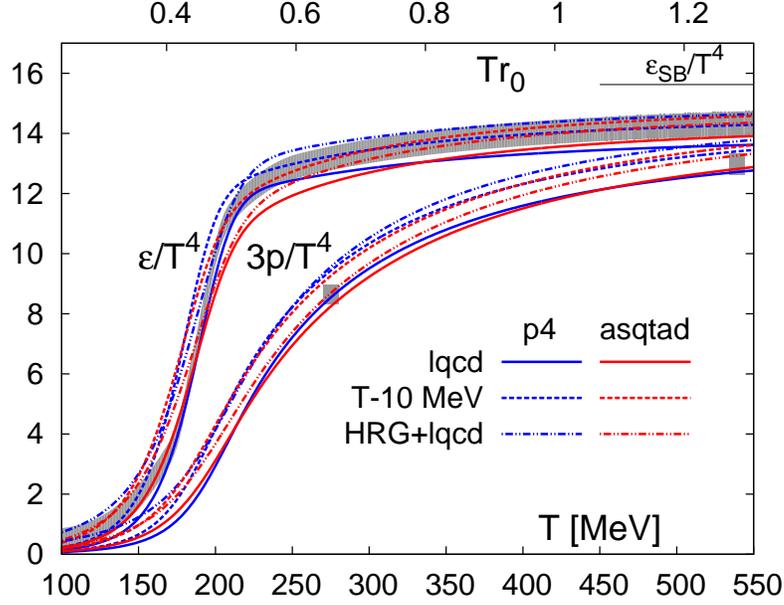,width=12.0cm}
\caption{(color online) Energy density and 3 times the pressure calculated
from new fitting Eq.~(\ref{eq:simplefit}) to the trace anomaly.  Fits to p4
and asqtad lattice data are solid lines, fits to lattice data shifted by
negative 10~MeV are dashed lines, and combined fits to HRG and lattice data
are double-dot dashed.  The systematic error band associated with beginning
the pressure integration at $T=100$~MeV and plotted as a black shaded box in
Fig.~\ref{fig:eos} is plotted as a grey band, with lower bound defined by
the pressure integration derived from the p4 interpolation.  The systematic
errors associated the interpolation are plotted here as grey boxes.  These
were shown as narrow error bars in Fig.~\ref{fig:eos}.}
\label{fig:e3p_2}
\end{center}
\end{figure}

Figure~\ref{fig:e3p_2} shows the energy density and pressure curves for  all fits compared to the systematic error calculations that were described in Section~\ref{sec:thermodynamics}.  These parametrized curves do not differ appreciably from the p4 and asqtad results shown in Fig.~\ref{fig:eos}, except that the asqtad equation of state has been extrapolated beyond the highest temperature data point at $\sim$400~MeV.  
Fits to the HRG merged to asqtad data lead to small increases in the
pressure but they are within the systematic errors associated with the
interpolation, shown as shaded boxes.  They shaded boxes were drawn as
narrow error bars in Fig.~\ref{fig:eos} that were centered on the p4
interpolation pressure curves.  The HRG+p4 merged and 10~MeV shifted data
fits lie slightly above this systematic, but fall below the systematic error
associated with using the HRG value for the pressure to begin the
integration at $T=100$~MeV, as plotted here as shaded band in the energy
density.  This error bar was shown
as a shaded box at high temperature in Fig.~\ref{fig:eos}.  The agreement
between the p4 and asqtad results at the highest temperature provides some
confidence in using the high temperature parametrization to extrapolate the
asqtad result up to 550~MeV.  At this temperature, all parametrizations are
below the Stefan-Boltzmann limit.
\begin{figure}[thb]
\begin{center}
\epsfig{file=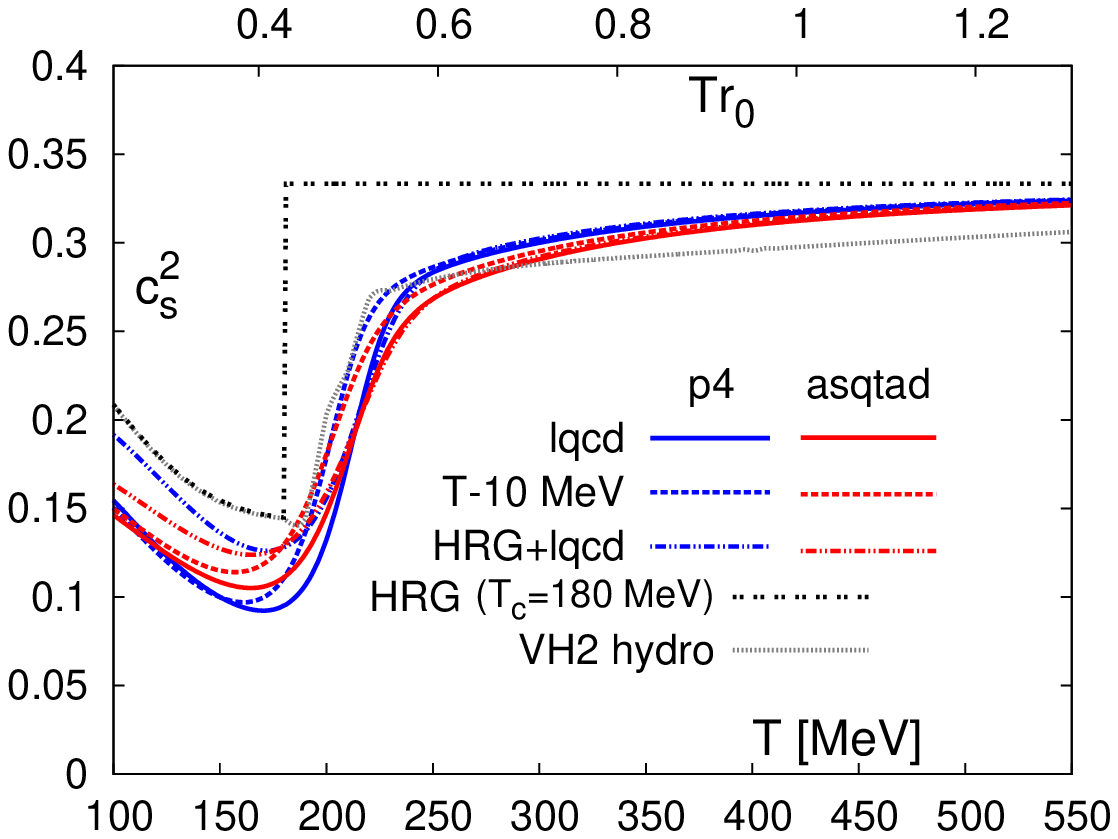,width=12.0cm}
\caption{(color online) Square of the velocity of sound, $c_s^2$, for the 
Eq.~(\ref{eq:simplefit}) 
parametrization of the trace anomaly for fits to the lattice data (solid) lattice 
data shifted by 10~MeV (dashed), and combined HRG and lattice calculations 
(double-dot dashed).  The lattice curves are compared to two typically EoS inputs 
currently used in hydrodynamic codes: the qcdEOS used in vh2 (dotted), and a HRG 
calculation with a first order transition at 180~MeV (double-dotted).}
\label{fig:cs2}
\end{center}
\end{figure}
 
 The square of the velocity of sound is shown in Fig.~\ref{fig:cs2}, as given by Eq.~(\ref{sound}).  Differences between the fits are mainly evident at lower temperatures.  The parametrizations for p4 and asqtad uniformly approach but do not attain the ideal gas limit shown for the HRG calculations with first order phase transition.  The latter is typical for many of the hydrodynamic calculations that are represented in the literature.  The EoS employed by the publicly available Viscous Hydro 2D+1 code (VH2)~\cite{Luzum:2008cw}  comes close to the set of lattice curves but falls somewhat below these new lattice results at higher temperatures.

\section{Summary of expectation values needed to calculate the trace anomaly}
\label{data_summary}

The various expectation values needed to evaluate the trace anomaly 
are summarized in Tables~\ref{tab:table_0} and \ref{tab:table_nz} for the
p4 action and in Tables~\ref{tab:asqtadTzeroGauge} to \ref{tab:asqtadNt6Fermion}  
for the asqtad action.

\begin{table}
\begin{tabular}{|c|c|c|c|c|c|c|c|}
\hline 
$\beta$  &  $m_l$ & $N_{\sigma}^3\times N_{\tau}$  & $\tau_{MD}$ & traj  & $\langle s_G \rangle_0$ & 
$\langle \bar \psi \psi \rangle_{l,0}$        & $\langle \bar \psi \psi \rangle_{s,0}$ \\ 
\hline 
 3.4300 & 0.00370 & $32^3\times 32$ & 0.5 & 3570 & 4.111679(148) & 0.075873(102) & 0.143475(74) \\ 
 3.4600 & 0.00313 & $32^3\times 32$ & 1.0 & 2060 & 4.044602(82) & 0.057641(85) & 0.117406(61) \\ 
 3.4900 & 0.00290 & $32^3\times 32$ & 0.5 & 2330 & 3.984160(149) & 0.044554(87) & 0.100634(57) \\ 
 3.5000 & 0.00253 & $32^3\times 32$ & 0.5 & 2150 & 3.964162(117) & 0.039426(93) & 0.089291(68) \\ 
 3.5100 & 0.00259 & $32^3\times 32$ & 0.5 & 1790 & 3.946336(118) & 0.036767(61) & 0.087441(51) \\ 
 3.5150 & 0.00240 & $32^3\times 32$ & 1.0 & 2100 & 3.936885(100) & 0.034644(64) & 0.081973(45) \\ 
 3.5225 & 0.00240 & $32^3\times 32$ & 1.0 & 2210 & 3.923687(111) & 0.032628(87) & 0.079799(56) \\ 
 3.5300 & 0.00240 & $32^3\times 32$ & 0.5 & 2870 & 3.910643(78) & 0.030849(38) & 0.077800(25) \\ 
 3.5400 & 0.00240 & $32^3\times 32$ & 0.5 & 3750 & 3.893451(86) & 0.028546(63) & 0.075221(48) \\ 
 3.5450 & 0.00215 & $32^3\times 32$ & 0.5 & 2270 & 3.884507(105) & 0.026400(77) & 0.068601(61) \\ 
 3.5500 & 0.00211 & $32^3\times 32$ & 0.5 & 2650 & 3.876406(101) & 0.025483(45) & 0.066852(31) \\ 
 3.5600 & 0.00205 & $32^3\times 32$ & 1.0 & 2090 & 3.859529(88) & 0.023064(85) & 0.063168(53) \\ 
 3.5700 & 0.00200 & $32^3\times 32$ & 0.5 & 1880 & 3.843775(75) & 0.021405(57) & 0.060291(36) \\ 
 3.5850 & 0.00192 & $32^3\times 32$ & 0.5 & 2310 & 3.820193(102) & 0.019054(55) & 0.056042(43) \\ 
 3.6000 & 0.00192 & $32^3\times 32$ & 0.5 & 3020 & 3.797184(65) & 0.017178(38) & 0.053636(28) \\ 
 3.6300 & 0.00170 & $24^3\times 32$ & 0.5 & 3232 & 3.752910(95) & 0.013176(93) & 0.045175(64) \\ 
 3.6600 & 0.00170 & $32^3\times 32$ & 0.5 & 2850 & 3.710875(72) & 0.011655(39) & 0.042370(27) \\ 
 3.6900 & 0.00150 & $24^3\times 32$ & 0.5 & 2284 & 3.669910(81) & 0.008740(85) & 0.035734(45) \\ 
 3.7600 & 0.00139 & $32^3\times 32$ & 0.5 & 3250 & 3.580229(48) & 0.006227(29) & 0.029547(12) \\ 
 3.8200 & 0.00125 & $32^3\times 32$ & 0.5 & 2430 & 3.508141(72) & 0.004507(69) & 0.024679(37) \\ 
 3.9200 & 0.00110 & $32^3\times 32$ & 0.5 & 4670 & 3.396463(44) & 0.002970(41) & 0.019633(9) \\ 
 4.0000 & 0.00092 & $32^3\times 32$ & 0.5 & 5430 & 3.313344(46) & 0.001788(28) & 0.015390(21) \\ 
 4.0800 & 0.00081 & $32^3\times 32$ & 0.5 & 5590 & 3.234959(32) & 0.001552(54) & 0.012778(19) \\ 
\hline 
\end{tabular} 
\caption{Parameters for simulations performed with the p4 action
at zero temperature and the expectation values 
of the gauge action and light and heavy
chiral condensates. Here we used also the zero temperature results from 
the study of the equation of state by RBC-Bielefeld
collaboration at $\beta=3.63$, 3.69, 3.82, and 3.92 \cite{rbcBIeos}. }
\label{tab:table_0}
\end{table}

\begin{table}
\begin{tabular}{|c|c|c|c|c|c|c|c|c|c|c|}
\hline 
$T$[MeV] & $\beta$  &  $m_l$ & $\tau_{MD}$ & traj  & $\langle s_G \rangle_\tau$ & $\langle \bar \psi \psi \rangle_{l,\tau}$ & 
$\langle \bar \psi \psi \rangle_{s,\tau}$ & $\Theta_G^{\mu\mu}/T^4$ & $\Theta_{F,l}^{\mu\mu}/T^4$ & $\Theta_{F,s}^{\mu\mu}/T^4$ \\ 
\hline 
    139 & 3.4300 & 0.00370 & 0.5 & 15140 & 4.111274(146) & 0.074251(111) & 0.142977(75) & 0.48(25) & 0.0433(42) & 0.133(28) \\ 
    154 & 3.4600 & 0.00313 & 0.5 & 11970 & 4.043850(110) & 0.055231(83) & 0.116653(58) & 0.95(18) & 0.0512(32) & 0.160(20) \\ 
    170 & 3.4900 & 0.00290 & 0.5 & 11800 & 3.983028(116) & 0.040290(103) & 0.099071(65) & 1.53(27) & 0.0758(30) & 0.278(17) \\ 
    175 & 3.5000 & 0.00253 & 1.0 & 10070 & 3.962857(99) & 0.034018(145) & 0.087194(75) & 1.81(23) & 0.0806(33) & 0.312(18) \\ 
    180 & 3.5100 & 0.00259 & 0.5 & 12660 & 3.944532(104) & 0.030218(98) & 0.084857(59) & 2.58(24) & 0.0958(21) & 0.378(13) \\ 
    183 & 3.5150 & 0.00240 & 1.0 & 10110 & 3.934529(79) & 0.026463(138) & 0.078531(66) & 3.41(20) & 0.1085(25) & 0.456(12) \\ 
    187 & 3.5225 & 0.00240 & 0.5 & 30620 & 3.920960(122) & 0.023138(237) & 0.075811(122) & 4.02(25) & 0.1218(34) & 0.512(17) \\ 
    191 & 3.5300 & 0.00240 & 0.5 & 43480 & 3.907753(92) & 0.020257(149) & 0.073347(81) & 4.34(18) & 0.1316(19) & 0.553(10) \\ 
    196 & 3.5400 & 0.00240 & 0.5 & 44880 & 3.890373(90) & 0.016457(137) & 0.070059(79) & 4.74(21) & 0.1439(23) & 0.614(13) \\ 
    199 & 3.5450 & 0.00215 & 0.5 & 13110 & 3.880942(187) & 0.012615(300) & 0.062299(198) & 5.55(37) & 0.1440(42) & 0.658(26) \\ 
    201 & 3.5500 & 0.00211 & 0.5 & 30140 & 3.872654(65) & 0.011318(93) & 0.060225(62) & 5.91(24) & 0.1423(22) & 0.666(12) \\ 
    206 & 3.5600 & 0.00205 & 1.0 & 9500 & 3.856090(118) & 0.008940(96) & 0.056184(87) & 5.54(30) & 0.1328(27) & 0.656(17) \\ 
    211 & 3.5700 & 0.00200 & 0.5 & 32580 & 3.840126(77) & 0.007533(94) & 0.052919(95) & 6.01(26) & 0.1228(27) & 0.653(18) \\ 
    219 & 3.5850 & 0.00192 & 1.0 & 4610 & 3.816785(77) & 0.005885(44) & 0.048085(83) & 5.80(31) & 0.1069(23) & 0.646(18) \\ 
    227 & 3.6000 & 0.00192 & 0.5 & 12750 & 3.794094(71) & 0.005337(16) & 0.046010(30) & 5.42(27) & 0.0925(20) & 0.596(14) \\ 
    243 & 3.6300 & 0.00170 & 0.5 & 12040 & 3.750217(76) & 0.004032(12) & 0.037784(35) & 4.99(32) & 0.0600(17) & 0.485(14) \\ 
    259 & 3.6600 & 0.00170 & 0.5 & 10610 & 3.708640(95) & 0.003694(6) & 0.035527(27) & 4.35(30) & 0.0507(10) & 0.436(9) \\ 
    275 & 3.6900 & 0.00150 & 0.5 & 14630 & 3.668111(64) & 0.003033(3) & 0.029784(17) & 3.66(25) & 0.0318(8) & 0.331(6) \\ 
    315 & 3.7600 & 0.00139 & 0.5 & 10740 & 3.578960(68) & 0.002536(1) & 0.025183(9) & 2.81(19) & 0.0192(2) & 0.228(1) \\ 
    351 & 3.8200 & 0.00125 & 0.5 & 15140 & 3.507192(39) & 0.002138(0) & 0.021318(6) & 2.23(22) & 0.0114(5) & 0.162(4) \\ 
    416 & 3.9200 & 0.00110 & 0.5 & 27180 & 3.395840(25) & 0.001743(0) & 0.017408(2) & 1.57(17) & 0.0054(3) & 0.098(3) \\ 
    475 & 4.0000 & 0.00092 & 0.5 & 23280 & 3.312885(38) & 0.001390(0) & 0.013887(1) & 1.21(19) & 0.0015(1) & 0.057(2) \\ 
    539 & 4.0800 & 0.00081 & 0.5 & 24140 & 3.234697(20) & 0.001175(0) & 0.011747(0) & 1.21(19) & 0.0015(1) & 0.057(2) \\ 
\hline 
\end{tabular} 
\caption{Parameters for simulations performed with the p4 action 
at finite temperatures, the expectation values of the gauge action, 
light and heavy chiral condensates, as well as the gluonic and fermionic
contributions to the trace anomaly obtained from them.}
\label{tab:table_nz}
\end{table}

\begin{table}[ht]
\begin{tabular}{|c|c|c|c|c|c|c|}
\hline
\maths{\beta}&\maths{m_l}&traj&\maths{u_0}&\maths{\VEV{P}}&\maths{\VEV{R}}&\maths{\VEV{C}} \\
\hline
6.400 & 0.00909 & 4830 & 0.8520 & 0.526573(7) & 0.271218(9)  & 0.277200(10) \\ 
6.430 & 0.00862 & 4935 & 0.8535 & 0.530550(7) & 0.276220(10) & 0.282561(10) \\ 
6.458 & 0.00820 & 6015 & 0.8549 & 0.534163(6) & 0.280792(8)  & 0.287447(9)  \\ 
6.500 & 0.00765 & 5390 & 0.8569 & 0.539353(5) & 0.287379(6)  & 0.294465(7)  \\ 
6.550 & 0.00705 & 5680 & 0.8594 & 0.545370(5) & 0.295095(7)  & 0.302650(7)  \\ 
6.600 & 0.00650 & 5215 & 0.8616 & 0.550963(6) & 0.302287(8)  & 0.310253(9)  \\ 
6.625 & 0.00624 & 4890 & 0.8626 & 0.553620(7) & 0.305709(10) & 0.313862(10) \\ 
6.650 & 0.00599 & 5310 & 0.8636 & 0.556230(5) & 0.309083(9)  & 0.317410(9)  \\ 
6.675 & 0.00575 & 5225 & 0.8647 & 0.558830(6) & 0.312461(9)  & 0.320960(9)  \\ 
6.700 & 0.00552 & 5055 & 0.8657 & 0.561350(4) & 0.315727(6)  & 0.324388(6)  \\ 
6.730 & 0.00525 & 5195 & 0.8668 & 0.564267(5) & 0.319526(7)  & 0.328364(8)  \\
6.760 & 0.00500 & 4922 & 0.8678 & 0.567073(7) & 0.323173(9)  & 0.332197(11) \\ 
6.800 & 0.00471 & 4755 & 0.8692 & 0.570777(4) & 0.328002(5)  & 0.337237(6)  \\ 
6.850 & 0.00437 & 4540 & 0.8709 & 0.575240(5) & 0.333853(7)  & 0.343340(8)  \\ 
6.900 & 0.00407 & 4310 & 0.8726 & 0.579580(4) & 0.339553(7)  & 0.349283(7)  \\ 
6.950 & 0.00380 & 4285 & 0.8741 & 0.583757(4) & 0.345060(6)  & 0.355003(7)  \\ 
7.000 & 0.00355 & 4130 & 0.8756 & 0.587820(6) & 0.350440(9)  & 0.360573(10) \\ 
7.080 & 0.00310 & 3965 & 0.8779 & 0.594080(4) & 0.358757(6)  & 0.369187(7)  \\ 
7.460 & 0.00180 &  251 & 0.8876 & 0.620817(4) & 0.394840(5)  & 0.406313(5)  \\
\hline
\end{tabular}
\caption{Simulation parameters at zero temperature for the asqtad
action and the expectation values of terms in the gauge action.  The
column labeled $P$ is the plaquette, $R$, the rectangle, and $C$, the
parallelogram. In all cases, the trajectory length is $\tau_{MD} = 1$
and the lattice size is $N_\sigma\times N_\tau = 32^4$, except for
$\beta = 7.46$, where it is $64^3144$ \label{tab:asqtadTzeroGauge}.}
\end{table}

\begin{table}[ht]
\begin{tabular}{|c|c|c|c|c|c|c|}
\hline
\maths{\beta} & \maths{\pbp_{l,0}} & \maths{\pbp_{s,0}} & \maths{\pbdmdup_{l,0}} & \maths{\pbdmdup_{s,0}} \\
\hline
6.400 & 0.101993(42) & 0.222407(30) & \maths{-4.89560(7)} & \maths{-4.82871(7)} \\
6.430 & 0.091265(62) & 0.207453(43) & \maths{-4.89481(7)} & \maths{-4.83519(9)} \\
6.458 & 0.082138(36) & 0.194171(26) & \maths{-4.89283(6)} & \maths{-4.83933(6)} \\
6.500 & 0.070319(33) & 0.176287(24) & \maths{-4.88921(5)} & \maths{-4.84355(5)} \\
6.550 & 0.058049(39) & 0.156758(28) & \maths{-4.88009(5)} & \maths{-4.84239(5)} \\
6.600 & 0.048335(44) & 0.139653(30) & \maths{-4.87296(5)} & \maths{-4.84183(5)} \\
6.625 & 0.044200(39) & 0.131874(31) & \maths{-4.87004(6)} & \maths{-4.84180(6)} \\
6.650 & 0.040376(40) & 0.124493(32) & \maths{-4.86661(4)} & \maths{-4.84097(4)} \\
6.675 & 0.036985(45) & 0.117575(29) & \maths{-4.86087(4)} & \maths{-4.83760(5)} \\
6.700 & 0.033917(39) & 0.111080(25) & \maths{-4.85648(4)} & \maths{-4.83535(4)} \\
6.730 & 0.030557(40) & 0.103681(30) & \maths{-4.85230(4)} & \maths{-4.83354(4)} \\
6.760 & 0.027679(30) & 0.096988(21) & \maths{-4.84924(7)} & \maths{-4.83249(6)} \\
6.800 & 0.024320(34) & 0.089120(23) & \maths{-4.84319(4)} & \maths{-4.82871(4)} \\
6.850 & 0.020852(67) & 0.080307(42) & \maths{-4.83519(4)} & \maths{-4.82318(5)} \\
6.900 & 0.017971(39) & 0.072717(28) & \maths{-4.82687(3)} & \maths{-4.81669(4)} \\
6.950 & 0.015605(43) & 0.066099(27) & \maths{-4.81963(3)} & \maths{-4.81103(4)} \\
7.000 & 0.013560(43) & 0.060154(26) & \maths{-4.81222(4)} & \maths{-4.80494(3)} \\
7.080 & 0.010803(49) & 0.050807(28) & \maths{-4.80044(3)} & \maths{-4.79507(3)} \\
7.460 & 0.0043819(81)& 0.0256237(100)&\maths{-4.74518(2)} & \maths{-4.74365(2)} \\
\hline
\end{tabular}
\caption{Continuation of the previous table.  Fermion expectation
  values for the asqtad action contributing to the equation of state.
  The last two columns give the contributions of asqtad gauge and
  fermion observables to the interaction measure.
\label{tab:asqtadTzeroFermion}
}
\end{table}

\begin{table}[ht]
\begin{tabular}{|c|c|c|c|c|c|c|c|}
\hline
\maths{T}[MeV]&\maths{\beta}&\maths{m_l}&traj&\maths{u_0}&\maths{\VEV{P}}&\maths{\VEV{R}}&\maths{\VEV{C}} \\
\hline
141 & 6.458 & 0.00820 & 14095 & 0.8549 & 0.534193(7) & 0.280840(10) & 0.287487(11) \\ 
149 & 6.500 & 0.00765 & 16943 & 0.8569 & 0.539410(7) & 0.287467(9)  & 0.294557(11) \\ 
160 & 6.550 & 0.00705 & 14605 & 0.8594 & 0.545430(7) & 0.295185(10) & 0.302728(11) \\ 
170 & 6.600 & 0.00650 & 14735 & 0.8616 & 0.551060(6) & 0.302443(9)  & 0.310395(10) \\ 
175 & 6.625 & 0.00624 & 16610 & 0.8626 & 0.553727(5) & 0.305883(8)  & 0.314021(9)  \\ 
181 & 6.650 & 0.00599 & 16235 & 0.8636 & 0.556370(6) & 0.309315(9)  & 0.317627(10) \\ 
183 & 6.658 & 0.00590 & 15655 & 0.8640 & 0.557250(7) & 0.310468(10) & 0.318830(11) \\ 
184 & 6.666 & 0.00583 & 16525 & 0.8643 & 0.558053(6) & 0.311505(9)  & 0.319920(9)  \\ 
186 & 6.675 & 0.00575 & 14900 & 0.8647 & 0.559000(6) & 0.312740(9)  & 0.321214(10) \\ 
188 & 6.683 & 0.00567 & 15010 & 0.8650 & 0.559823(8) & 0.313814(13) & 0.322341(15) \\ 
190 & 6.691 & 0.00560 & 14695 & 0.8653 & 0.560617(6) & 0.314844(10) & 0.323429(11) \\ 
192 & 6.700 & 0.00552 & 13735 & 0.8657 & 0.561547(6) & 0.316064(8)  & 0.324702(9)  \\ 
194 & 6.708 & 0.00544 & 10695 & 0.8659 & 0.562290(7) & 0.317019(10) & 0.325701(10) \\ 
195 & 6.715 & 0.00538 & 11690 & 0.8662 & 0.563000(7) & 0.317951(10) & 0.326673(12) \\ 
199 & 6.730 & 0.00525 & 12690 & 0.8668 & 0.564490(6) & 0.319902(9)  & 0.328716(10) \\ 
202 & 6.745 & 0.00512 & 10900 & 0.8673 & 0.565900(7) & 0.321732(11) & 0.330624(11) \\ 
206 & 6.760 & 0.00500 & 14815 & 0.8678 & 0.567293(5) & 0.323553(8)  & 0.332525(9)  \\ 
215 & 6.800 & 0.00471 & 14185 & 0.8692 & 0.570997(8) & 0.328391(11) & 0.337577(10) \\ 
227 & 6.850 & 0.00437 & 14035 & 0.8709 & 0.575460(5) & 0.334240(8)  & 0.343663(9)  \\ 
240 & 6.900 & 0.00407 & 14295 & 0.8726 & 0.579770(5) & 0.339900(7)  & 0.349547(8)  \\ 
252 & 6.950 & 0.00380 & 14270 & 0.8741 & 0.583933(7) & 0.345387(10) & 0.355247(12) \\ 
266 & 7.000 & 0.00355 & 14460 & 0.8756 & 0.587977(5) & 0.350733(6)  & 0.360787(8)  \\ 
288 & 7.080 & 0.00310 & 14595 & 0.8779 & 0.594227(4) & 0.359030(6)  & 0.369370(7)  \\ 
409 & 7.460 & 0.00180 &  3415 & 0.8876 & 0.620880(3) & 0.394973(4)  & 0.406340(5)  \\ 
\hline
\end{tabular}
\caption{Simulation parameters at nonzero temperature for the asqtad
action and the expectation values of terms in the gauge action.  The
column labeled $P$ is the plaquette, $R$, the rectangle, and $C$, the
parallelogram. In all cases the trajectory length is $\tau_{MD} = 1$ and 
the lattice size is $N_\sigma^3\times N_\tau = 32^3\times 8$, except at 
$\beta = 7.46$, where it is $64^3\times8$.
\label{tab:asqtadThighGauge}
}
\end{table}

\begin{table}[ht]
\begin{tabular}{|c|c|c|c|c|c|c|c|}
\hline
\maths{\beta} & \maths{\pbp_{l,\tau}} & \maths{\pbp_{s,\tau}} & \maths{\pbdmdup_{l,\tau}} & \maths{\pbdmdup_{s,\tau}} &
\maths{\theta^{\mu\mu}_G/T^4} & \maths{\theta^{\mu\mu}_{F,l}/T^4} & \maths{\theta^{\mu\mu}_{F,s}/T^4} \\
\hline
6.458 & 0.079427(65)  & 0.193628(44)  & \maths{-4.89320(9)  } & \maths{-4.83982(7)  } & 0.50(14) & 0.1192(20) & 0.1192(65) \\
6.500 & 0.066841(58)  & 0.175448(37)  & \maths{-4.88992(7)  } & \maths{-4.84433(7)  } & 1.08(13) & 0.1339(15) & 0.1614(56) \\
6.550 & 0.053589(65)  & 0.155654(39)  & \maths{-4.88082(7)  } & \maths{-4.84324(6)  } & 1.08(13) & 0.1630(15) & 0.2015(52) \\
6.600 & 0.041831(95)  & 0.137746(48)  & \maths{-4.87408(7)  } & \maths{-4.84315(7)  } & 1.82(13) & 0.2257(15) & 0.3314(50) \\
6.625 & 0.036566(91)  & 0.129519(47)  & \maths{-4.87126(5)  } & \maths{-4.84332(6)  } & 2.00(13) & 0.2585(15) & 0.3985(47) \\
6.650 & 0.031162(150) & 0.121454(68)  & \maths{-4.86818(6)  } & \maths{-4.84294(8)  } & 2.70(13) & 0.3039(18) & 0.5014(48) \\
6.658 & 0.028996(110) & 0.118580(58)  & \maths{-4.86581(6)  } & \maths{-4.84147(7)  } & $-$      & $-$        & $-$        \\
6.666 & 0.027696(140) & 0.116377(61)  & \maths{-4.86482(6)  } & \maths{-4.84121(6)  } & $-$      & $-$        & $-$        \\
6.675 & 0.025791(130) & 0.113677(59)  & \maths{-4.86268(5)  } & \maths{-4.84001(6)  } & 3.26(13) & 0.3604(18) & 0.6271(48) \\
6.683 & 0.023944(170) & 0.111080(90)  & \maths{-4.86189(7)  } & \maths{-4.83998(9)  } & $-$      & $-$        & $-$        \\
6.691 & 0.022785(150) & 0.108953(72)  & \maths{-4.86079(6)  } & \maths{-4.83957(6)  } & $-$      & $-$        & $-$        \\
6.700 & 0.021084(120) & 0.106292(59)  & \maths{-4.85859(5)  } & \maths{-4.83817(5)  } & 3.91(13) & 0.4022(18) & 0.7512(48) \\
6.708 & 0.019759(160) & 0.103973(85)  & \maths{-4.85910(6)  } & \maths{-4.83941(7)  } & $-$      & $-$        & $-$        \\
6.715 & 0.018723(130) & 0.102063(64)  & \maths{-4.85745(6)  } & \maths{-4.83838(7)  } & $-$      & $-$        & $-$        \\
6.730 & 0.016480(100) & 0.097952(58)  & \maths{-4.85464(6)  } & \maths{-4.83669(6)  } & 4.48(14) & 0.4280(24) & 0.8704(42) \\
6.745 & 0.014737(95)  & 0.094147(69)  & \maths{-4.85306(6)  } & \maths{-4.83631(7)  } & $-$      & $-$        & $-$        \\
6.760 & 0.013427(74)  & 0.090633(61)  & \maths{-4.85140(5)  } & \maths{-4.83563(5)  } & 4.42(13) & 0.4198(9)  & 0.9355(39) \\
6.800 & 0.010537(49)  & 0.082012(52)  & \maths{-4.84533(7)  } & \maths{-4.83186(6)  } & 4.55(13) & 0.3592(6)  & 0.9257(33) \\
6.850 & 0.008404(26)  & 0.072736(40)  & \maths{-4.83720(4)  } & \maths{-4.82624(4)  } & 4.58(13) & 0.3088(6)  & 0.9392(29) \\
6.900 & 0.007124(21)  & 0.065144(36)  & \maths{-4.82857(4)  } & \maths{-4.81946(4)  } & 4.00(13) & 0.2560(4)  & 0.8929(20) \\
6.950 & 0.006207(10)  & 0.058684(28)  & \maths{-4.82111(5)  } & \maths{-4.81348(5)  } & 3.80(13) & 0.2105(4)  & 0.8303(17) \\
7.000 & 0.005496(6)   & 0.053087(16)  & \maths{-4.81346(4)  } & \maths{-4.80704(3)  } & 3.39(13) & 0.1708(4)  & 0.7500(15) \\
7.080 & 0.004517(4)   & 0.044388(10)  & \maths{-4.80151(3)  } & \maths{-4.79684(3)  } & 3.22(13) & 0.1217(4)  & 0.6222(13) \\
7.460 & 0.0022535(3)  & 0.0224922(11) & \maths{-4.745490(17)} & \maths{-4.744170(19)} & 1.68(10) & 0.0258(1)  & 0.1884(4)  \\
\hline
\end{tabular}
\caption{Continuation of the previous table.  Fermion expectation
  values for the asqtad action contributing to the $N_\tau = 8$
  equation of state.
  The last three columns give the contributions of asqtad gauge and
  fermion observables to the interaction measure.  Where a matching $T = 0$ point
  was not simulated, the entry is omitted.
\label{tab:asqtadThighFermion}
}
\end{table}

\begin{table}[ht]
\begin{tabular}{|c|c|c|c|c|c|c|c|}
\hline
\maths{T}[MeV]&\maths{\beta}&\maths{m_l}&traj&\maths{u_0}&\maths{\VEV{P}}&\maths{\VEV{R}}&\maths{\VEV{C}} \\
\hline
174 & 6.400 & 0.00909 & 18665 & 0.8520 & 0.526870(13) & 0.271659(18) & 0.277621(18) \\
181 & 6.430 & 0.00862 & 19090 & 0.8535 & 0.530953(11) & 0.276821(15) & 0.283137(15) \\
188 & 6.458 & 0.00820 & 19760 & 0.8549 & 0.534690(8)  & 0.281597(12) & 0.288216(13) \\
194 & 6.480 & 0.00791 & 17960 & 0.8560 & 0.537580(10) & 0.285303(15) & 0.292145(17) \\
199 & 6.500 & 0.00765 & 18285 & 0.8570 & 0.540170(12) & 0.288646(18) & 0.295674(19) \\
202 & 6.512 & 0.00750 &  7730 & 0.8575 & 0.541610(17) & 0.290493(24) & 0.297617(29) \\
206 & 6.525 & 0.00735 & 17805 & 0.8581 & 0.543193(7)  & 0.292536(10) & 0.299787(11) \\
213 & 6.550 & 0.00705 & 18250 & 0.8593 & 0.546153(8)  & 0.296357(12) & 0.303793(14) \\
227 & 6.600 & 0.00650 & 10367 & 0.8616 & 0.551790(8)  & 0.303634(12) & 0.311446(14) \\
234 & 6.625 & 0.00624 & 19305 & 0.8626 & 0.554420(6)  & 0.307026(9)  & 0.315005(9)  \\
248 & 6.675 & 0.00575 & 20565 & 0.8647 & 0.559547(6)  & 0.313662(9)  & 0.321958(9)  \\
256 & 6.700 & 0.00552 & 20420 & 0.8657 & 0.562003(7)  & 0.316846(10) & 0.325289(11) \\
\hline
\end{tabular}
\caption{Simulation parameters at nonzero temperature for the asqtad
action and the expectation values of terms in the gauge action.  The
column labeled $P$ is the plaquette, $R$, the rectangle, and $C$, the
parallelogram. In all cases the lattice size is $N_\sigma^3\times N_\tau 
= 32^3\times 6$ and the trajectory length is $\tau_{MD} = 1$.
\label{tab:asqtadNt6Gauge}
}
\end{table}

\begin{table}[ht]
\begin{tabular}{|c|c|c|c|c|c|c|c|}
\hline
\maths{\beta} & \maths{\pbp_{l,\tau}} & \maths{\pbp_{s,\tau}} & \maths{\pbdmdup_{l,\tau}} & \maths{\pbdmdup_{s,\tau}} &
\maths{\theta^{\mu\mu}_G/T^4} & \maths{\theta^{\mu\mu}_{F,l}T^4} & \maths{\theta^{\mu\mu}_{F,s}/T^4} \\
\hline
6.400 & 0.089082(160) & 0.218269(81)  & \maths{-4.89921(12)} & \maths{-4.83308(15)} & 1.65(5) & 0.1931(9)  & 0.3095(27) \\
6.430 & 0.074353(130) & 0.201725(62)  & \maths{-4.89961(12)} & \maths{-4.84093(10)} & 2.24(5) & 0.2425(9)  & 0.4108(27) \\
6.458 & 0.060070(150) & 0.186297(69)  & \maths{-4.89895(8) } & \maths{-4.84686(10)} & 2.96(4) & 0.3056(10) & 0.5451(25) \\
6.480 & 0.048998(140) & 0.174662(79)  & \maths{-4.89769(11)} & \maths{-4.85023(12)} & $-$     & $-$        & $-$        \\
6.500 & 0.039520(210) & 0.164169(110) & \maths{-4.89608(11)} & \maths{-4.85267(13)} & 4.75(4) & 0.3753(8)  & 0.7382(24) \\
6.512 & 0.035186(280) & 0.158502(160) & \maths{-4.89601(18)} & \maths{-4.85479(21)} & $-$     & $-$        & $-$        \\
6.525 & 0.030473(99)  & 0.152353(66)  & \maths{-4.89488(8) } & \maths{-4.85601(9) } & $-$     & $-$        & $-$        \\
6.550 & 0.024137(120) & 0.141436(88)  & \maths{-4.89066(7) } & \maths{-4.85595(9) } & 4.55(4) & 0.3920(5)  & 0.8856(21) \\
6.600 & 0.016492(50)  & 0.122340(61)  & \maths{-4.88119(8) } & \maths{-4.85347(8) } & 4.93(5) & 0.3499(4)  & 0.9513(20) \\
6.625 & 0.014284(25)  & 0.114105(44)  & \maths{-4.87786(5) } & \maths{-4.85304(6) } & 4.80(4) & 0.3205(3)  & 0.9520(15) \\
6.675 & 0.011338(12)  & 0.099656(29)  & \maths{-4.86748(4) } & \maths{-4.84763(4) } & 4.39(4) & 0.2611(3)  & 0.9123(12) \\
6.700 & 0.010336(9)   & 0.093466(27)  & \maths{-4.86232(4) } & \maths{-4.84447(5) } & 4.06(4) & 0.2341(3)  & 0.8742(11) \\
\hline
\end{tabular}
\caption{Continuation of the previous table.  Fermion expectation
  values for the asqtad action contributing to the $N_\tau = 6$ equation of state.
  The last three columns give the contributions of asqtad gauge and
  fermion observables to the interaction measure. Where a matching $T = 0$ point
  was not simulated, the entry is omitted.
\label{tab:asqtadNt6Fermion}
}
\end{table}

\end{document}